\documentclass[journal]{IEEEtran}

\ifCLASSINFOpdf
\else
\fi
\usepackage{scalefnt}
\usepackage[shortlabels]{enumitem}
\usepackage{graphicx}
\usepackage{amssymb}
\usepackage{amsmath}
\usepackage{amsfonts}
\usepackage{scalefnt}
\usepackage{color}
\usepackage{cite}
\usepackage{multicol}
\usepackage{theorem}
\usepackage{hyperref}
\usepackage{amsmath}
\usepackage{comment}
\usepackage{float}
\usepackage[caption=false]{subfig}
\usepackage{algorithm}
\usepackage{algorithmic}
\usepackage[normalem]{ulem}
\usepackage[dvipsnames]{xcolor}

\begin{document}
\scalefont{0.9789}
\bstctlcite{IEEEexample:BSTcontrol}
\title{A Framework for Holistic KLD-based Waveform Design for Multi-User-Multi-Target ISAC Systems}
\author{Yousef Kloob,~\IEEEmembership{Member,
~IEEE}, Mohammad Al-Jarrah,~\IEEEmembership{Member,
~IEEE}, and Emad Alsusa,~\IEEEmembership{Senior Member,~IEEE,}\vspace{-0.4in}\thanks{%
Y. Kloob, M. Al-Jarrah and E. Alsusa are with the Department of Electrical and
Electronic Engineering, University of Manchester, Manchester M13 9PL, U.K.
(e-mail: \{yousef.kloob, mohammad.al-jarrah, e.alsusa\}@manchester.ac.uk).}\thanks{%
}}

\maketitle

\begin{abstract}
This paper introduces a novel framework aimed at designing integrated waveforms for robust integrated sensing and communication (ISAC) systems. The system model consists of a multiple-input multiple-output (MIMO) base station that simultaneously serves communication user equipments (UEs) and detects multiple targets using a shared-antenna deployment scenario. By leveraging Kullback-Leibler divergence (KLD) to holistically characterise both communication and sensing subsystems, three optimisation problems are formulated: (i) radar waveform KLD maximisation under communication constraints, (ii) communication waveform KLD maximisation subject to radar KLD requirements, and (iii) an integrated waveform KLD-based optimisation for ISAC that jointly balances both subsystems. The first two problems are solved using a projected gradient method with adaptive penalties for the radar waveforms and a gradient-assisted interior point method (IPM) for the communication waveforms. The third, integrated waveform optimisation approach adopts an alternating direction method of multipliers (ADMM) framework to unify radar and communication waveform designs into a single integrated optimisation, thereby synergising sensing and communication objectives and achieving higher overall performance than either radar- or communication-only techniques. Unlike most existing ISAC waveform designs that regard communication signals solely as interference for sensing, the proposed framework utilises the holistic ISAC waveform---that is, the superimposed communication and sensing signals---to boost detection performance in the radar subsystem. Simulation results show significant improvements in both radar detection and communication reliability compared with conventional zero-forcing beamforming, identity-covariance radar baselines, and traditional optimisation approaches, demonstrating the promise of KLD-based waveform designs for next-generation ISAC networks.

\end{abstract}


\IEEEpeerreviewmaketitle
\vspace{-0.2 in}
\section{Introduction}
\IEEEPARstart{T}{he} rapid evolution of wireless communication networks has dramatically expanded the landscape of interconnected devices to include new applications such as autonomous vehicles and unmanned aerial vehicles (UAVs) that are heavily dependent on advanced sensing, pushing the boundaries of communication systems beyond their traditional limits \cite{7355569,9509294,9426946,9598915}. As we approach the sixth-generation (6G) era, network operators and service providers aim not only to further enhance their communication services but also to include advanced sensing-based applications such as detection, localisation, tracking, navigation, and environmental surveillance \cite{Ericsson,rajatheva2020white,9861699}.

Integrated Sensing and Communications (ISAC) systems have gained significant interest as a key enabler of many 6G applications. ISAC aims to synergistically merge sensing and communication functionalities, hence optimising base station (BS) resources for dual purposes \cite{9705498,7898445,8386661,9226446,9385108,9540344}. This integration can be achieved through various approaches. For example, in a separated deployment, the BS antennas are allocated distinctly between communication and radar subsystems, which can simplify the system design but may underutilise resources. In contrast, the shared deployment strategy employs all antennas for both functionalities, hence offering higher resource efficiency, but at the cost of increased complexity due to the need for more sophisticated signal processing techniques\cite{8288677,9200993}. 

The evolution of ISAC has been accelerated by advances in multi-antenna technologies, particularly massive Multiple-Input Multiple-Output (MIMO) systems, which significantly enhance both communication capacity and sensing accuracy \cite{Fatema2018MassiveML}. However, evaluating the performance of ISAC systems poses unique challenges due to their dual functionality. Traditionally, communication and sensing subsystems have been assessed using distinct metrics—communication performance through achievable rate, outage probability, and bit error rate (BER), and sensing performance through estimation rate, detection probability, false alarm probability, and mean square error (MSE) \cite{9800940}. This disparity complicates the holistic assessment and optimisation of ISAC systems. To this end, some unified performance measures have been introduced in the literature, such as mutual information \cite{10.5555/1146355,7086341,4776572,Fan0}, and Kullback-Leibler divergence
(KLD), also known as the relative entropy, which we recently introduced \cite{kloob,Al-Jarrah2023,kloob1,10118838}. Unlike mutual information approaches that assume capacity-achieving codes and focus on information transmission \cite{Shannon1948,5452208}, KLD directly measures distinguishability for practical finite-length systems, offers direct performance insights through Stein's lemma for hypothesis testing, and captures the impact of practical modulation schemes on system performance, aligning naturally with ISAC's dual objective.
\vspace{-0.12 in}
\subsection{Literature review}
\vspace{-0.02 in}
While the application of KLD in sensing systems is well-established \cite{7086341,5467189}, its potential in characterising communication system performance has recently emerged as a powerful tool for ISAC design. For instance, in our previous works \cite{Al-Jarrah2023,10118838}, we employed KLD to analyse the performance trade-off in ISAC systems with separated antenna deployment, relating radar detection probability $P_D$ and communication BER to the achievable KLD. This connection stems from the fact that KLD is intrinsically tied to the maximum-likelihood detection, widely used in radar detection and equally central to communication receivers, whose errors are quantified by BER. Consequently, KLD can quantify the detection capability at both the radar and communication ends in a similar manner, making it a powerful and unified design tool for ISAC systems. While in \cite{kloob} we proposed a low-complexity unified objective function based on KLD, the aim was focused on optimising network resources for both separated and shared deployment scenarios. Moreover, further insights about the achievable KLD trade-off in ISAC systems are explored in \cite{10636778}. Nonetheless, the literature lacks a comprehensive investigation and beamforming design for a generalised multi-user-multi-target ISAC system using the promising KLD framework. 
The fundamental strength of KLD lies in its ability to provide a unified measure that captures the key detection processes in both subsystems. Unlike traditional metrics that operate in disparate mathematical spaces—probability of detection for radar and BER for communication, KLD enables direct comparison and joint optimisation. Moreover, KLD offers analytical tractability through closed-form gradient expressions, avoiding the computational complexity associated with Q-functions and error function derivatives in traditional formulations.

Further efforts have explored various optimisation approaches for ISAC systems, each addressing specific aspects of system performance. For example, joint radar-communication beamforming optimisation based on the Cramér-Rao Bound (CRB) has been proposed to improve both sensing and communication performance \cite{9652071}, where this approach minimises the CRB of angle estimation while maximising the signal-to-interference-plus-noise ratio (SINR) for communication. However, these methods often focus on optimising individual components rather than adopting a unified framework that captures the interplay between different functionalities. Beamforming optimisation has been a particular focus, with studies investigating partially connected hybrid designs to address hardware complexity issues \cite{9868348}. While these approaches offer some practical advantages in terms of implementation, they fail to fully capture the nuanced trade-offs between sensing and communication performance in deeply integrated scenarios. Similarly, waveform optimisation has also been extensively studied, with recent work using faster-than-Nyquist approaches to enhance spectral efficiency while maintaining radar functionality as a constraint \cite{10298105}. Recent studies have also investigated joint transmit and receive beamforming in full-duplex ISAC systems to manage interference and enhance overall performance; however, these approaches primarily emphasise interference suppression rather than exploitation, thereby neglecting potential performance gains achievable through direct integration of communication signals into radar sensing processes \cite{10159012}. Despite these advances, existing work has not adequately addressed the need for a unified optimisation framework that simultaneously serves both subsystems’ requirements and balances their performance trade-offs effectively \cite{10158711}. Furthermore, a recent comprehensive survey on interference management in ISAC systems highlights that most current strategies remain oriented toward interference mitigation or cancellation, reinforcing the perspective of communication signals as detrimental rather than beneficial \cite{10770127}. Consequently, such designs not only overlook opportunities to leverage inherent signal interactions but also risk introducing excessive complexity, which is especially problematic when real-time processing is required in dynamic environments.
\vspace{-0.11 in}
\subsection{Motivation and contribution}
\vspace{-0.02 in}
Despite these advances, current approaches often fall short of exploiting the full potential of the integrated system by assuming sensing and communication are competing subsystems rather than considering a synergic design. Most notably, there is a need for more comprehensive models that accurately reflect the real-world operation of ISAC systems, particularly in scenarios where communication signals can be utilised for radar target detection alongside dedicated radar signals. Exploiting the interference from communication signals as a source of information for radar operations, as shown in our paper, enhances radar detection capabilities, improving the system's overall resource efficiency. This dual use of communication signals represents a significant departure from traditional approaches, which typically treat interference as a detrimental factor to be mitigated rather than leveraged\cite{7746569,8114253}.
Toward these objectives and motivated by the fact that KLD is able to provide a unified measure for ISAC systems holistically, this paper proposes a novel ISAC waveform design with the objective of maximising KLD for the whole system as a design criterion in a shared antenna deployment scenario. Additionally, this work investigates the synergies between sensing and communication in ISAC systems utilising KLD through well-established relations between KLD on one side and the detection probability in radar via Stein’s Lemma and BER in communication systems from the other side \cite{Al-Jarrah2023}. Our system model consists of a MIMO BS that simultaneously serves communication user equipment (UEs) and detects multiple targets using shared resources, using a statistical radar model. To this end, the key contributions of this work are:
\begin{itemize}[leftmargin=1em]  
\item Two Novel Separate Waveform Optimisation Designs: we present radar and communication waveform optimisation using the unified KLD measure to represent the achievable performance for both subsystems on one scale. Both formulated problems aim at maximising the KLD measure of a certain subsystem subject to the minimum KLD of the other subsystem and power constraints. The non-convex radar optimisation employs a projected gradient method with an adaptive penalty, while the communication optimisation uses a gradient-assisted interior point method. These techniques efficiently handle ISAC waveform design's complex nature, enabling balanced performance optimisation across both subsystems within system constraints.

\item Integrated Waveform Optimisation for ISAC: Beyond the individual radar and communication waveform-oriented designs, an integrated waveform optimisation framework for ISAC is introduced to optimise both waveforms together. This integrated waveform approach leverages an alternating direction method of multipliers (ADMM) algorithm, yielding enhanced performance in radar detection accuracy and communication reliability compared with the separate waveform methods. By capturing and exploiting the synergy between sensing and data transmission, the unified approach achieves superior performance in shared-antenna ISAC deployments.

\item Exploitation of Communication Interference: Our framework accounts for the fact that communication signals can also be exploited for radar target detection alongside dedicated radar signals, providing a more comprehensive and useful model of ISAC system operation. This dual-use approach enhances the radar performance by utilising communication interference as an additional source of information for target detection, which offers new insights into the underlying synergy between radar and communication waveforms and quantifies how their joint optimisation can enhance overall ISAC capabilities.

\item Complexity Evaluation: We provide a thorough computational complexity analysis of our proposed ISAC algorithms, evaluating their scalability and real-time feasibility. This analysis is critical for determining practical implementation strategies in dynamic network environments where swift adaptation and efficient resource utilisation are essential.

\item Benchmark Analysis:  We compare the proposed KLD-based strategies against the conventional zero-forcing (ZF) beamforming for the communication subsystem and the identity covariance design for the radar subsystem. In addition, we introduce two optimisation frameworks based on the BER and probability of detection to further benchmark our methods. These comparisons highlight the performance gains, computational complexity trade-offs, and practical relevance of our KLD-based approach in a range of ISAC scenarios.

\end{itemize}
\vspace{-0.04 in}
Our results demonstrate that the proposed KLD-based optimisation techniques significantly outperform ZF beamforming and the radar subsystem's identity covariance baseline, as well as the BER-based and detection-based designs in terms of computational cost. The radar waveform KLD-based optimisation achieves substantial gains in both target detection and communication quality, whereas the communication waveform KLD-based optimisation primarily benefits the communication subsystem, yielding modest gains in radar performance. Both separate waveform techniques exhibit robust performance across varying SNR levels, with the radar waveform design demonstrating notably stable computational efficiency. The integrated waveform optimisation KLD-based for ISAC approach unifies both radar and communication waveforms into a single integrated design, delivering higher overall performance in terms of sensing and communication signals. KLD-based formulations also facilitate a structured, scalable, and computationally efficient optimisation framework—unlike complex BER-based and detection-based designs, as evidenced by our numerical results. Notably, utilising the communication signals in radar detection significantly enhances overall system performance, especially in scenarios with limited dedicated radar resources. These highlight the practical significance of KLD-based waveform design and provide key insights for developing efficient shared deployment ISAC systems that balance sensing and communication requirements.

\vspace{-0.12 in}
\subsection{Paper organisation}
The paper is structured as follows. Sec.II presents the system model. Sec.III and IV analyse the communication and radar systems, respectively, including the derivations of KLD for each system. In Sec.V, the radar waveform optimisation is investigated. Sec.VI, the communication waveform optimisation is investigated. Sec.VII, introduces the integrated waveform optimisation. Sec.VIII shows the complexity analysis for all waveform optimisation techniques, Sec.IX displays the numerical results, and finally, Sec.X concludes the work.

\textit{$Notation$}: Bold uppercase letters (e.g., $\mathbf{S}$) denote matrices, and bold lowercase letters (e.g., $\mathbf{s}$) denote vectors. Superscripts ${(\cdot)}^*$, ${(\cdot)}^T$, and ${(\cdot)}^H$, denote the conjugate, transpose, and Hermitian transpose, respectively. Subscripts ${(\cdot)}_\mathrm{c}$ and ${(\cdot)}_\mathrm{r}$ relate to the communication, and radar subsystems, respectively. 
\vspace{-0.07 in}
\section{System Model}

We consider an $N$ antenna MIMO-BS. The antennas are utilised for detecting a maximum number of ${T}$ targets and serving $K$ number of single-antenna communication UEs in the downlink direction in a shared deployment manner, reflecting the emerging paradigm where next-generation wireless infrastructure will inherently support dual-functional operation for both communication and sensing services \cite{9737357}. The total transmitted power available at BS is $P_\mathrm{T}$ which is utilised for both sensing and data communication duties. The power $P_\mathrm{T}$ allocated to the radar and communication subsystems are denoted as $P_\mathrm{r}$ and $P_\mathrm{c}$, where $P_\mathrm{T}=P_\mathrm{c}+P_\mathrm{r}$. As a starting point, a non-optimised ZF beamforming technique is employed at BS to precode the information of communication UEs \cite{Fatema2018MassiveML}. These techniques serve as benchmarks for our subsequent optimised systems, which will be introduced and analysed later in Sections V and VI. The combined transmitted ISAC signal $\mathbf{x}_l\in \mathbb{C}^{N\times 1}$ at the $l$-th snapshot can be formulated as follows,
\vspace{-0.05 in}
\begin{equation}
    \mathbf{x}_l=\mathbf{W}_{\mathrm{c}}\mathbf{s}_{\mathrm{c},l}+\mathbf{W}_{\mathrm{r},l}\mathbf{s}_{\mathrm{r}},\vspace{-0.05 in}
\end{equation}
where $l \in \{1,2,\dots,L\}$ represents the discrete time index, and $L$ is the total number of snapshots considered for sensing outcome, $\mathbf{s}_{\mathrm{c},l}\in \mathbb{C}^{K\times 1}$ is a vector of communication UEs symbols, $\mathbf{s}_{\mathrm{r}}\in \mathbb{C}^{T\times 1}$ is a vector of the baseband radar waveforms for the potential targets, $\mathbf{W}_{\mathrm{c}}\in \mathbb{C}^{N\times K}$ represents the precoding matrix for the communication subsystem, and $\mathbf{W}_{\mathrm{r},l}\in \mathbb{C}^{N\times T}$ represents the precoding matrix for the radar subsystem. The transmitted signals vector $\mathbf{x}_l$ is transmitted from the BS and received at the UEs as well as the targets, which reflect the received $\mathbf{x}_l$ back to the BS. Although the radar signal component received at the UEs might cause interference that cannot be eliminated by the users, the communication signal component received at the targets and reflected back to the BS carries information about the target. Therefore, by carefully designing the precoding matrices at the BS, the sensing information can be enhanced by exploiting the communication signal. This dual-use approach enhances the radar performance by utilising the communication signal as an additional source of information. The radar waveform design is based on the combined transmit signal covariance matrix for the $t$-th beam where the covariance matrix of the $t$-th transmit beam is $\mathbf{R}_{t}=\frac{1}{L}\sum_{l=1}^{L}\mathbf{x}_{t,l}\:\mathbf{x}_{t,l}^H$, $\mathbf{x}_{t,l}=\mathbf{w}_{\mathrm{r},t,l}s_{\mathrm{r},t}+\mathbf{W}_{\mathrm{c}}\mathbf{s}_{\mathrm{c},l}$. This formulation allows us to incorporate the communication signals into the radar detection process. It should be noticed that the power for the radar is integrated into the precoding vector $\mathbf{w}_{\mathrm{r},t,l}$, where it is designed to have $\| \mathbf{w}_{\mathrm{r},t,l} \|_F^2 = \frac{P_\mathrm{r}}{T}$, and $\mathbf{w}_{\mathrm{r},t,l}=\frac{P_\mathrm{r}}{NT}\:\mathbf{\tilde{w}}_{\mathrm{r},t,l}$, where $\mathbf{\tilde{w}}_{\mathrm{r},t,l}$ is the $t$th beam precoding vector, and $\|.\|_F$ is the Frobenius norm.
\vspace{-0.12 in}
\subsection{Communication System}\label{sec-ii-a}
At each $l$ instance, a data symbol $s_{\mathrm{c},k,l}$ intended for the $k$-th UE is drawn from a normalised constellation, i.e, $\mathbb{E}[ \left\vert s_{\mathrm{c},k,l} \right\vert ^{2}] =1$. The received signal at the $k$-th UE can be formulated as 
\vspace{-0.05 in}
\begin{align}
y_{\mathrm{c},k,l}=\overbrace{\mathbf{h}_{k}^T\:d_{\mathrm{c},k}^{-\zeta/2}\:\mathbf{w}_{\mathrm{c},k}s_{\mathrm{c},k,l}}^\text{Desired $k$-th UE signal}+\overbrace{\omega_{\mathrm{MN},k}}^{\substack{\text{Inter-user}\\ \text{interference}}} +\overbrace{\eta_k}^{\substack{\text{Radar interference}\\ \text{and noise}}},\label{1}
\end{align}
where $\mathbf{h}_{k}\in \mathbb{C}^{N\times 1}\sim \mathcal{C}\mathcal{N}\left( 0,\sigma _{h}^{2}\right) $ represents the channel from the BS to the $k$-th UE, $\omega_{\mathrm{MN},k}=\sum_{i=1,i \neq k}^{K}{\mathbf{h}_{k}^T\:d_{\mathrm{c},k}^{-\zeta/2}\:\mathbf{w}_{\mathrm{c},i}s_{\mathrm{c},i,l}}$ represents the IUI on the $k$-th UE from the other UEs, $d_{\mathrm{c},k}^{-\zeta/2}$ is the channel pathloss from BS to the $k$-th UE with $d_{\mathrm{c},k}$ representing the distance from BS to the $k$-th UE, and $\zeta$ is the pathloss exponent $\eta_k=\mathbf{h}_{k}^T\:d_{\mathrm{c},k}^{-\zeta/2}\:\mathbf{W}_{\mathrm{r},l}\mathbf{s}_{\mathrm{r},l} +n_{k,l}$ is the interference-plus-noise term with ${n}_{k,l}\sim \mathcal{C}\mathcal{N}\left( 0,\:\sigma _{n}^{2}\right) $ representing the additive white Gaussian noise (AWGN). The precoding vector for the communication system is $\mathbf{w}_{\mathrm{c},k}\in \mathbb{C}^{N\times 1}$, which is typically designed based on the given channel matrix from MIMO-BS to UEs $\mathbf{H} \in \mathbb{C}^{N\times K}= \left[\mathbf{h}_{1},.. ,\mathbf{h}_{k},.., \mathbf{h}_{K} \right]$, where the elements of $\mathbf{H}$ are independent and identically distributed (i.i.d) complex Gaussian random variables with zero mean and variance $\sigma_h^2$. We assume that channel state information (CSI) is obtained through standard pilot-based estimation procedures widely adopted in modern communication systems with established techniques such as MMSE and ML estimators \cite{1597555,6457363}. 

\vspace{-0.15 in}
\subsection{Radar system}
The radar system is designed for high adaptability, allowing real-time adjustments in detection capabilities across $\!L\!$ frame snapshots. $T\!$ denotes the radar beams emitted during a given detection frame. Using MIMO radar technology, multiple beams can be formed simultaneously via orthogonal signals \cite{7126203, 5419124}, where $T$ is the upper limit of detectable targets in one frame.

The system explores multiple angular-range-Doppler bins over successive frames, increasing target identification. Its flexibility in varying emitted beams allows for the simultaneous detection of more targets, to fine-tune the system’s detection capability according to different operational demands. However, it is crucial to acknowledge that the value of $T$ is constrained by the available antennas and UEs serviced by the BS, highlighting the need for efficient waveform design and resource allocation in ISAC systems. Our focus aligns with scenarios where targets are spatially separated, with each target confined to a distinct radar bin, as per prior studies \cite{9119156,9229166} \footnote{The literature presents algorithms to separate signals from closely spaced or unresolved targets, facilitating accurate target enumeration \cite{8964303}.}.
The total radar return signal at the $l$-th snapshot is shown as follows,
\vspace{-0.07 in}
\begin{equation} 
\mathbf{y}_{\mathrm{r},l}=\sum_{t=1}^{T}{\mathbf{H}_{t}^T\:d_{\mathrm{r},t}^{\zeta/2}\:\mathbf{w}_{\mathrm{r},t,l}s_{\mathrm{r},t}}+\mathbf{H}_{t}^T\:d_{\mathrm{r},t}^{\zeta/2}\:\mathbf{W}_{\mathrm{c}}\mathbf{s}_{\mathrm{c},l}+\mathbf{n}_{l},\vspace{-0.05 in}
\end{equation}
where $\mathbf{H}_t \in \mathbb{C}^{N \times N}$ is the target response matrix that captures the complete BS-Target-BS path characteristics for the $t$-th target, inherently incorporating propagation effects, target cross-section, and reflection coefficients. The target response matrix follows a Rayleigh distribution to model the composite effects of multiple scattering centres inherent in extended targets, thereby characterising the statistical nature of target scattering while preserving the direct path essential for target detection \cite{1597550,4408448}. The \( \mathbf{n}_{l} \sim \mathcal{CN}(\mathbf{0}, \sigma_n^2\mathbf{I}_N) \) is the AWGN vector at the \( l \)-th snapshot. 
To process the radar returns from all beams, a bank of matched filters is typically applied, each tuned to a specific waveform $s_{\mathrm{r},t}: \forall t=\left\{ 1,2,\cdots ,T \right\}$, which corresponds to a particular radar angular-range-Doppler bin. Assuming $s_{\mathrm{r},t}$ is orthogonal to $s_{\mathrm{r},i}$ for all $t \neq i$ using signal space and time domain, the radar returns from distinct beams can be effectively separated, enabling independent detection of each potential target \cite{4408448,5419124}. The radar precoding matrix $\mathbf{W}_{\mathrm{r},l}$ is designed such that the covariance matrix is $\mathbf{R}_{w}\triangleq \frac{1}{L}\sum_{l=1}^{L}\mathbf{W}_{\mathrm{r},l}\mathbf{W}_{\mathrm{r},l}^{H}\in \mathbb{C}^{N\times N}$. The received signal for the \( t \)-th target at the BS after match-filtering, under the binary hypothesis testing problem \( \mathcal{H}_{q}\forall q \in \{0,1\} \), where \( \mathcal{H}_{1} \) represents the target presence whereas \( \mathcal{H}_{0} \) denotes the target absence, is denoted by $\mathbf{y}_{\mathrm{r},t,l|\mathcal{H}_{q}}$ and can be mathematically represented as follows,
\vspace{-0.05 in}
\label{r35}
\begin{equation} 
\mathbf{y}_{\mathrm{r},t,l|\mathcal{H}_{q}}=\overbrace{\mathbf{H}_{t}^T\:d_{\mathrm{r},t}^{\zeta/2}\:\mathbf{w}_{\mathrm{r},t,l}}^{\substack{\text{Radar waveform}\\ \text{ radar return}}}\:q+\overbrace{\mathbf{H}_{t}^T\:d_{\mathrm{r},t}^{\zeta/2}\:\mathbf{W}_{\mathrm{c}}\mathbf{s}_{\mathrm{c},l}}^{\substack{\text{Comms. waveform}\\ \text{ radar return}}}q+\overbrace{\mathbf{n}_{t,l}}^{\text{Noise}},\label{e-y-t}
\end{equation}
where \( \mathbf{n}_{t,l} \sim \mathcal{CN}(\mathbf{0}, \sigma_n^2\mathbf{I}_N) \) is the AWGN vector at the \( l \)-th snapshot at the output of the matched filter. 
\vspace{-0.1 in}
\section{The communication KLD: ZF and Arbitrary precoding matrix \label{sec:CommKLD}}

In this section, we derive the KLD of the communication system for ZF beamforming with shared deployment antenna configuration, motivated by its direct connection to the maximum likelihood detection process and tractable analytical expressions for system optimisation, where the KLD quantifies the distinguishability between different transmitted symbols by measuring the information divergence of their received signal distributions, capturing the fundamental ability to differentiate communication symbols at the receiver. The analysis is carried out for the normalised precoders introduced in Section \ref{sec-ii-a}. To derive the KLD, it is necessary to establish the statistical properties of the signal, IUI, radar interference, and noise. For a pair of multivariate Gaussian distributed random variables having mean vectors of $\mathbf{\mu}_m$ and $\mathbf{\mu}_n$ and covariance matrices of $\Sigma_m$ and $\Sigma_n$, the KLD can be derived as,
\vspace{-0.06 in}
\begin{multline}
\mathrm{KLD}_{n\rightarrow m}=\frac{1}{2\ln 2}\left( \mathrm{tr}\left(
\Sigma _{n}^{-1}\Sigma _{m}\right) -2+\left( \mathbf{\mu }_{n}-\mathbf{\mu 
}_{m}\right) ^{T} \right. \\
\; \left.  \times\;\Sigma _{n}^{-1} \left( \mathbf{\mu }_{n}-\mathbf{\mu }%
_{m}\right) +\ln \frac{\left\vert \Sigma _{n}\right\vert }{\left\vert
\Sigma _{m}\right\vert }\right). \label{zb}
\end{multline}
\vspace{-0.3 in}
\subsection{ZF beamforming}
ZF designs are typically used to precode the communication data symbols at the BS. This scheme will serve as a benchmark for our designs. ZF eliminates inter-user interference (IUI) at the UEs regardless of the amount of the received power. This usually results in excellent performance at high SNR values. The normalised ZF beamforming matrix is also designed based on the channel matrix, and can be represented as follows,
\vspace{-0.05 in}
\begin{equation}
\mathbf{W}_{\mathrm{c}}=\mathbf{P}\mathbf{\tilde{W}}_{\mathrm{c}}/\|\mathbf{\tilde{W}}_{\mathrm{c}} \|_F
\end{equation}
where $\mathbf{\tilde{W}}_{\mathrm{c}}=\mathbf{H}^{H}\:\left(\mathbf{H}\:\mathbf{H}^{H}\right)^{-1}$is the ZF beamformer before normalisation, and $\mathbf{P}\in \mathbb{C}^{K\times K}=\mathrm{diag}\left( \sqrt{p_{1}},\cdots ,\sqrt{p_{K}}\right)$ is the power allocation matrix for the communication users, which is a diagonal matrix that controls the power allocated for each UE. With ZF precoding, the received signal at the $k$-th user is,
\vspace{-0.03 in}
\begin{equation}
y_{k,l} = \mathbf{h}_{k}^Td_{\mathrm{c},k}^{-\zeta/2}\mathbf{w}_{\mathrm{c},k}s_{\mathrm{c},k} + \mathbf{h}_{k}^T d_{\mathrm{c},k}^{-\zeta/2}\mathbf{W}_{\mathrm{r},l}\mathbf{s}_{\mathrm{r},l} + n_{k,l},
\end{equation}
where $\mathbf{w}_{\mathrm{c},k}$ is the $k$-th column of $\mathbf{W}_{\mathrm{c}}$.  It is worth noting that the term $\sum_{i=1,i \neq k}^{K}{\mathbf{h}_{k}^T\:d_{\mathrm{c},k}^{-\zeta/2}\:\mathbf{w}_{\mathrm{c},i}s_{\mathrm{c},i,l}}=0$ as the ZF precoder eliminates the IUI. The radar interference term $\mathbf{h}_{k}^T d_{\mathrm{c},k}^{-\zeta/2}\mathbf{W}_{\mathrm{r},l}\mathbf{s}_{\mathrm{r},l}$ consists of $NT$ independent random variables arising from the random radar precoding. For moderate to large $NT$ values, CLT ensures that this sum converges to a Gaussian distribution. Combined with the inherently Gaussian noise $n_{k,l}$, the total interference-plus-noise $\eta_k$ is well-approximated as $\mathcal{CN}(0, \sigma_\eta^2)$, particularly for large $N$. This approximation yields a mean of $\mathbb{E}[\eta_k]=0$, and a variance of
\vspace{-0.03 in}
\begin{align}
\sigma_\eta^2 &=d_{\mathrm{c},k}^{-\zeta} \mathbb{E}[\|\mathbf{h}_{k}^T \mathbf{W}_{\mathrm{r},l}\|^2] + \sigma_n^2 = d_{\mathrm{c},k}^{-\zeta}\sigma_h^2\mathrm{Tr}(\mathbf{R}_{w})+\sigma_n^2
\end{align}
The mean vectors and covariance matrices for the $n$-th and $m$-th symbols are given by,
\vspace{-0.06 in}
\begin{align}
\boldsymbol{\mu}_{k,j} &= d_{\mathrm{c},k}^{-\zeta/2} \mathbb{E}[\mathbf{h}_{k}^T \mathbf{w}_{\mathrm{c},k}] s_{\mathrm{c},k,l}^{(j)} \nonumber \\
&= d_{\mathrm{c},k}^{-\zeta/2} \mathbb{E}\left[\|\mathbf{\tilde{W}}_{\mathrm{c}} \|_F^{-2}\right] s_{\mathrm{c},k,l}^{(j)}, \quad j \in \{n, m\}, \label{uzf}\\\
\Sigma_n &= \Sigma_m = \sigma_\eta^2,\label{sigzf}
\end{align}
where $\mathbb{E}[\mathbf{h}_{k}^T\mathbf{w}_{\mathrm{c},k}] = 1$ due to the normalisation of the ZF precoding matrix. Since $\|\mathbf{\tilde{W}}_{\mathrm{c}} \|_F^{-2}=\mathrm{tr}((\mathbf{H} \mathbf{H}^H)^{-1})$, under flat fading channel, $\|\mathbf{\tilde{W}}_{\mathrm{c}} \|_F^{-2}$ follows a Gamma distribution with shape parameter $N-K+1$ and scale parameter $1$, denoted as $\|\mathbf{\tilde{W}}_{\mathrm{c}} \|_F^{-2} \sim \mathrm{Gamma}(L_{\mathrm{G}}, 1)$, where $L_{\mathrm{G}} = N-K+1$. Let $x \triangleq \|\mathbf{\tilde{W}}_{\mathrm{c}}\|_F^2$. By substituting \eqref{uzf}, and \eqref{sigzf} in \eqref{zb}, and considering all possible pairs of dissimilar symbols \cite[Corollary 1]{Al-Jarrah2023}, then the KLD becomes,
\vspace{-0.05 in}
\begin{equation}
\mathrm{KLD}_{\mathrm{c},k}^{\mathrm{ZF}} = \frac{\lambda \:d_{\mathrm{c},k}^{-\zeta}\:p_k}{M(M-1) \sigma_\eta^2 \ln 2} \mathbb{E}\left[\frac{1}{\|\mathbf{\tilde{W}}_{\mathrm{c}}\|_F^2}\right] \label{zf1},
\end{equation}
where $\lambda =\sum_{m=1}^{M}\sum_{\substack{ n=1  \\ n\neq m}}%
^{M}\left\vert s_{\mathrm{c},k,l}^{(n)} -s_{\mathrm{c},k,l}^{(m)} \right\vert
^{2}$ is a constant that depends on the constellation. The expectation of $1/\|\mathbf{\tilde{W}}_{\mathrm{c}}\|_F^2$ can be calculated using the PDF of Gamma distribution as follows,
\vspace{-0.03 in}
\begin{align}
\begin{split}
\mathbb{E}\!\left[\|\mathbf{\tilde{W}}_{\!\mathrm{c}}\|_F^{-2}\right] \!\!&=\!\!\! \int_0^\infty\!\! \frac{1}{x} \cdot \frac{1}{\Gamma(L_{\mathrm{G}})} x^{L_{\mathrm{G}}-1} e^{-x}\! dx \!=\! N\!\!-\!\!K.
\end{split}
\end{align}
\vspace{-0.02 in}
Substituting this result into \eqref{zf1}, we get the final $\mathrm{KLD}_{\mathrm{c},k}^{\mathrm{ZF}}$,
\vspace{-0.03 in}
\begin{equation}
\mathrm{KLD}_{\mathrm{c},k}^{\mathrm{ZF}} = \lambda\:d_{\mathrm{c},k}^{-\zeta}\: p_k (N-K)/M(M-1) \sigma_\eta^2 \ln 2.
\end{equation}
\subsection{Conditional KLD for an Arbitrary Precoding Matrix \texorpdfstring{$\mathbf{W}_\mathrm{c}$}{Wc} and Given Channel Matrix}
A conditional KLD for communication, in terms of the beamforming matrix $\mathbf{W}_\mathrm{c}$ and for a given channel matrix $\mathbf{H}$, is required to formulate an optimisation problem. By noting that the received signal representation $y_{\mathrm{c},k,l}$ in \eqref{1} consists of a desired term, $(K-1)$ statistically independent inter-user symbols and $NT$ independent radar/noise terms, as well as, by considering CLT, the resulting sum is well approximated by a proper complex Gaussian for a typical MIMO-ISAC system which employs large antenna arrays.  We therefore take $y_{\mathrm{c},k,l}^{(j)}\sim\mathcal{CN}(\mu_{k,j},\Sigma_{\mathrm c,j}),\ \ j \in \{n, m\}$, where the conditional mean vectors and variance are,
\vspace{-0.03 in}
\begin{equation}
\mu_{k,j} = \mathbb{E}[y_{\mathrm{c},k,l}^{(j)}] = \mathbf{h}_{k}^T\:d_{\mathrm{c},k}^{-\zeta/2}\:\mathbf{w}_{\mathrm{c},k}s_{\mathrm{c},k,l}^{(j)}, \!\quad j \in \{n, m\},\label{mu1}\\
\end{equation}
\begin{align}
\operatorname{Var}(y_{\mathrm{c},k,l}^{(j)}) &= \mathbb{E}[|y_{\mathrm{c},k,l}^{(j)} - \boldsymbol{\mu}_{k,n}|^2] \nonumber \\
&=\! d_{\mathrm{c},k}^{-\zeta}\!\!\!\sum_{i=1,i\neq k}^{K}\!\!\!\|\mathbf{h}_{k}^T\mathbf{w}_{\!\mathrm{c},i}\|^2 \!+\! \sigma_{\eta}^2, \!\!\!\quad j \in \{n, m\}.\label{sig1}
\end{align}
Note the variance of the received signal for a given pair of symbols $\{s_{\mathrm{c},k,l}^{(n)},s_{\mathrm{c},k,l}^{(m)}\}\forall\{m,n\},m\neq n$ is equal. Thus, the covariance is,
\vspace{-0.08 in}
\begin{equation}\Sigma_{\mathrm{c},n} = \Sigma_{\mathrm{c},m} = d_{\mathrm{c},k}^{-\zeta}\:\sum_{i=1,i\neq k}^{K}\|\mathbf{h}_{k}^T\mathbf{w}_{\mathrm{c},i}\|^2 +  \sigma_{\eta}^2,\label{sig2}\end{equation} 
by substituting the evaluated mean vectors and covariance matrices in \eqref{mu1}, and \eqref{sig2} into the KLD expression given in \eqref{zb}, we can obtain the KLD for the $k$-th UE in the communication subsystem for each possible pair of unequal data symbols $\{s_{\mathrm{c},k,l}^{(n)}, s_{\mathrm{c},k,l}^{(m)}\}$, $n \neq m$, which can be represented as follows,
\vspace{-0.03 in}
\begin{align}
&\!\!\mathrm{KLD}^{m\rightarrow n}_{\mathrm{c},k} \!=\!\frac{d_{\mathrm{c},k}^{-\zeta}}{2\ln 2}\Bigg(\!\!\left(\mathbf{h}_{k}^T\mathbf{w}_{\mathrm{c},k}(s_{\mathrm{c},k,l}^{(m)} \!- \!s_{\mathrm{c},k,l}^{(n)})\right)^H\nonumber\\
 &\!\!\times\! \Bigg(\!d_{\mathrm{c},k}^{-\zeta}\!\!\!\sum_{i=1,i\neq k}^{K}\!\!\!\!\|\mathbf{h}_{k}^T\mathbf{w}_{\mathrm{c},i}\|^2 \!+\!  \sigma_{\eta}^2\Bigg)^{\!\!\!-1} 
\!\!\!\!\!\! \left(\mathbf{h}_{k}^T\mathbf{w}_{\mathrm{c},k}(s_{\mathrm{c},k,l}^{(m)} \!- \!s_{\mathrm{c},k,l}^{(n)})\!\right)\!\!\Bigg)\!.\!\!\label{e-kld-d-d}
\end{align}
Thereafter, by considering all possible pairs of dissimilar symbols, the KLD expression using $\lambda$ is simplified as,
\vspace{-0.03 in}
\begin{equation}
\begin{split}
\mathrm{KLD}^{m\rightarrow n}_{\mathrm{c},k} &=\frac{\lambda\:d_{\mathrm{c},k}^{-\zeta}}{2  M(M-1)\ln 2} \!\!\times\!\! \frac{\|\mathbf{h}_{k}^T\mathbf{w}_{\mathrm{c},k}\|^2}{d_{\mathrm{c},k}^{-\zeta}\!\sum_{i=1,i\neq k}^{K}\!\|\mathbf{h}_{k}^T\mathbf{w}_{\mathrm{c},i}\|^2 \!+\! \sigma_{\eta}^2}.
\end{split}
\end{equation}
Where the interference is the sum of IUI and radar interference. It should be noted that the second term in this equation represents a form of the SINR, confirming KLD's alignment with conventional metrics while uniquely enabling modulation-aware distinguishability and synergistic interference exploitation through its unified hypothesis-testing foundation.
\vspace{-0.05 in}
\section{Radar system analysis \label{RAD-ANA}}
\subsection{Composite detection and response matrix estimation}
The main objective of radar systems is to detect the existence of targets and estimate the matrix response for existing targets. After separating signals associated with different targets, we formulate a composite detection-estimation problem as follows,
\vspace{-0.05 in}
\begin{equation}
\{\hat{\mathbf{H}}_t,\hat{\mathcal{H}}_q\} = \arg\max_{\mathbf{H}_t,\mathcal{H}_q\forall q\in\{0,1\} }f(\mathbf{y}_{\mathrm{r},t,l};\mathbf{H}_t,\mathbf{x}_{t,l},\mathcal{H}_{q}).
\end{equation}
This joint optimisation problem aims to simultaneously detect the presence of a target, i.e., determining $\hat{\mathcal{H}}_q$, and estimate its response matrix $\hat{\mathbf{H}}_t$ if present. 
It can be observed from (\ref{e-y-t}) that the distribution of the received signals vector \( \mathbf{y}_{\mathrm{r},t,l} \) under each hypothesis is multivariate normal whose PDF is,
\vspace{-0.05 in}
\begin{align}
f(\mathbf{y}_{\mathrm{r},t,l};\, & \mathbf{H}_t q,\, \mathbf{x}_{t,l} q,\, \mathcal{H}_q) = 
\frac{1}{\pi^{N} \left[ (1 - q)\sigma_n^{2N} + q\, \det(\mathbf{R}_{1,t}) \right]} \nonumber \\
&\times \exp \left( -\mathbf{y}_{\mathrm{r},t,l}^H \left[ (1 - q)\sigma_n^{-2} 
+ q\, \det(\mathbf{R}_{1,t}) \right] \mathbf{y}_{\mathrm{r},t,l} \right), \label{fq}
\end{align}
where \( \mathbf{R}_{1,t}=\mathbf{H}_t \mathbf{R}_{t,l} \mathbf{H}_t^H+\sigma_n^2\mathbf{I}_N \), and \( \mathbf{R}_{t,l}= \mathbf{x}_{t,l}\:\mathbf{x}_{t,l}^H\:d_{\mathrm{r},t}^{\zeta} \). The generalised likelihood ratio test (GLRT) for the \( t \)-th target is used to determine target existence or absence by comparing the likelihoods under the two hypotheses, which is derived as,
\vspace{-0.05 in}
\begin{equation}
\Lambda_{t,l} = \frac{\arg\max_{\mathbf{H}_t} f(\mathbf{y}_{\mathrm{r},t,l};\mathbf{H}_t,\mathbf{x}_{t,l},\mathcal{H}_{1})}{\arg\max_{\mathbf{H}_t} f(\mathbf{y}_{\mathrm{r},t,l};\mathcal{H}_{0})}.
\end{equation}
As the PDF does not depend on $\mathbf{H}_t$ under $\mathcal{H}_0$, the target response matrix is estimated under $\mathcal{H}_1$, thus by using the maximum likelihood estimation theorem, we obtain,
\vspace{-0.06 in}
\begin{equation}
\hat{\mathbf{H}}_t = \arg\max_{\mathbf{H}_t} f(\mathbf{y}_{\mathrm{r},t,l};\mathbf{H}_t,\mathbf{x}_{t,l},\mathcal{H}_{1}).
\label{est}
\end{equation}
Since the PDF of $\mathbf{y}_{\mathrm{r},t,l}$ is multivariate Gaussian, the maximisation leads to the least squares solution given by,
\vspace{-0.03 in}
\begin{align}
\hat{\mathbf{H}}_t \!&=\! \arg\min_{\mathbf{H}_t} \left\|\mathbf{y}_{\mathrm{r},t} \!- \!\mathbf{H}_t\mathbf{x}_{t,l}\right\|^2 \!=\!\mathbf{y}_{\mathrm{r},t}\mathbf{x}_{t,l}^H\left(\mathbf{x}_{t,l}\mathbf{x}_{t,l}^H\right)^{-1}.
\end{align} 
This least squares solution provides an unbiased estimate of the target‐response matrix $\mathbf{H}_t$, with accuracy directly tied to the detection probability $P_D$. When the radar correctly detects a target under $\mathcal{H}_1$, the maximum likelihood solution $\hat{\mathbf{H}}_t$ is used to estimate $\mathbf{H}_t$. In contrast, a miss detection (i.e., deciding $\mathcal{H}_0$ under $\mathcal{H}_1$) forces $\hat{\mathbf{H}}_t=\mathbf{0}$, resulting in significant errors. The overall unconditional MSE is defined as,
\vspace{-0.03 in}
\begin{align}
\mathrm{MSE} &= p(H_1)\Bigl[(1-P_D)\,\mathrm{MSE}_{0\_1} + P_D\,\mathrm{MSE}_{1\_1}\Bigr] \nonumber\\&+ \bigl[1-p(H_1)\bigr]\Bigl[P_{\mathrm{FA}}\,\mathrm{MSE}_{1\_0}\Bigr],
\label{eq:MSE_overall}
\end{align}
where $p(H_1)$ is the prior probability of target existence with equal probability assumed, $\mathrm{MSE}_{1\_1}\triangleq\mathbb{E}[\|\hat{\mathbf{H}}_t-\mathbf{H}_t\|^2\mid \mathcal{H}_q=\mathcal{H}_1,\hat {\mathcal{H}}_q=\mathcal{H}_1]$ is the MSE under correct detection, $\mathrm{MSE}_{1\_0}\triangleq\mathbb{E}[\|\hat{\mathbf{H}}_t\|^2\mid \mathcal{H}_q=\mathcal{H}_0,\hat{\mathcal{H}}_q=\mathcal{H}_1]$ is the MSE for a false detection, and $\mathrm{MSE}_{0\_1}\triangleq\mathbb{E}[\|\hat{\mathbf{H}}_t-\mathbf{H}_t\|^2\mid \mathcal{H}_q=\mathcal{H}_1,\hat{\mathcal{H}}_q=\mathcal{H}_0]$ is the MSE for a miss detection (with $\hat{\mathbf{H}}_t=\mathbf{0}$) \footnote{$\mathrm{MSE}_{0\_0}\triangleq 0$ is not used because when the target is absent and correctly declared absent, both the true and estimated $\mathbf{H}_t$ are zero, yielding no error.}. Evaluating these expectations yields $\mathrm{MSE}_{1\_1} = \mathrm{MSE}_{1\_0} = \sigma_n^2\,\frac{N}{P_\mathrm{r}}$ and $\mathrm{MSE}_{0\_1} = N^2\sigma_n^2$. Thus, improving $P_D$ not only increases the frequency of valid estimations but also mitigates the impact of miss detections, thereby reducing the overall error and reinforcing how enhanced detection and lowering false alarms directly improve estimation accuracy.\\ 
Substituting the PDF in \eqref{fq} into the GLRT, we obtain,
\vspace{-0.07 in}
\begin{align}
&\Lambda(\mathbf{y}_{\mathrm{r},t,l})= \nonumber \\
& \frac{\!\sigma_n^{2N} \!\exp\!\left(\!-(\mathbf{y}_{\mathrm{r},t,l}\! -\! \hat{\mathbf{H}}_t\mathbf{x}_{t,l})^H\hat{\mathbf{R}}_{1,t}^{-1}(\mathbf{y}_{\mathrm{r},t,l} \!-\! \hat{\mathbf{H}}_t\mathbf{x}_{t,l})\!\right)}{ \!\det(\hat{\mathbf{R}}_{1,t})\exp\!\left(\!-\frac{1}{\sigma_n^2}\mathbf{y}_{\mathrm{r},t,l}^H\:\mathbf{y}_{\mathrm{r},t,l}\!\right)}\!\!\underset{\mathcal{H}_0}{\overset{\mathcal{H}_1}{\gtrless}} \!\tau_0
\end{align}
where $\hat{\mathbf{R}}_{1,t} = \hat{\mathbf{H}}_t\mathbf{R}_{t,l}\hat{\mathbf{H}}_t^H + \sigma_n^2\mathbf{I}_N$ and $\tau_0$ is the detection threshold. Taking the logarithm of the likelihood ratio and rearranging terms, the final GLRT form is given by,
\begin{equation} 
\dot{\Lambda}(\mathbf{y}_{\mathrm{r},t,l}) = \mathbf{y}_{\mathrm{r},t,l}^H\left(\mathbf{I}_N+\hat{\mathbf{R}}_{1,t}^{-1}\right)\mathbf{y}_{\mathrm{r},t,l}\underset{\mathcal{H}_0}{\overset{\mathcal{H}_1}{\gtrless}} \tau_1\label{e-tozta5}
\end{equation}  
where  $\tau_1 = \ln \tau_0 + N\ln\sigma_n^2 - \ln\det(\hat{\mathbf{R}}_{1,t})$. Here, the constant terms \(N\ln\sigma_n^2\) and \(\ln\det(\hat{\mathbf{R}}_{1,t})\) have been absorbed into the threshold. This expression decides the most probable hypothesis, i.e., \( \mathcal{H}_0 \) or \( \mathcal{H}_1 \), by incorporating the estimated target response matrix \( \hat{\mathbf{H}}_t \) in the covariance matrix \( \mathbf{R}_{1,t} \). Since the received signals from different snapshots are conditionally independent, (\ref{e-tozta5}) can be updated to adapt to $L$ snapshot scenario by taking the average, i.e., $\bar{\Lambda}_{t,l} = \frac{1}{L} \sum_{l=1}^{L} \dot{\Lambda}_{t,l}\underset{\mathcal{H}_0}{\overset{\mathcal{H}_1}{\gtrless}} \tau_2$, with $\tau_2$ is the detection threshold considering all snapshots and can be designed according to Neyman-Pearson lemma.
\vspace{-0.05 in}
\subsection{The KLD of radar system \label{sec:kldrad}}

As the number of snapshots $L$ approaches infinity, the law of large numbers ensures that the estimated target response matrix converges in probability to the true value, i.e., $\hat{\mathbf{H}}_t \xrightarrow{p} \mathbf{H}_t$. This allows us to use $\mathbf{H}_t$ in place of $\hat{\mathbf{H}}_t$ for large $L$. The received radar signal $\mathbf{y}_{r,t,l}$ becomes a linear transformation of the Gaussian noise $\mathbf{n}_{t,l}$ plus deterministic signal components, where we can proceed with a more tractable derivation of the KLD for the radar subsystem.

Using this property, we can now derive the $\mathrm{KLD}_{\mathrm{r},t}^{\mathcal{H}_0\rightarrow \mathcal{H}_1}$ for the radar subsystem using the probability density functions established in \eqref{fq}, and \eqref{zb} as follows,
\vspace{-0.03 in}
\begin{align}
\mathrm{KLD}_{\mathrm{r},t}^{\mathcal{H}_0\rightarrow \mathcal{H}_1}=& \frac{1}{\ln{2}}\Big(\ln\left(\mathrm{det}(\mathbf{R}_{2,t})\right)+\mathrm{Tr}\left(\mathbf{R}_{2,t}^{-1}\:\left(\sigma_n^2\:\mathbf{I}_N\right)\right) \nonumber\\
&-N\:\left(1+\ln\left(\sigma_n^2\right)\right)\Big),\label{kldrt}
\end{align}
where $\mathbf{R}_{2,t}=\frac{1}{L}\sum_{l=1}^{L}\mathbf{R}_{1,t}= \mathbf{H}_t\frac{1}{L}\sum_{l=1}^{L}\mathbf{R}_{t,l}\mathbf{H}_t^H+\sigma_n^2\mathbf{I}_N=\mathbf{H}_t\mathbf{R}_t\mathbf{H}_t^H+\sigma_n^2\mathbf{I}_N$, and $\mathrm{det}(.)$ is the determinant operation. $\mathrm{KLD}_{\mathrm{r},t}=\frac{1}{2}\left(\mathrm{KLD}_{\mathrm{r},t}^{\mathcal{H}_0\rightarrow \mathcal{H}_1}+\mathrm{KLD}_{\mathrm{r},t}^{\mathcal{H}_1\rightarrow \mathcal{H}_0}\right)$, where in this case $\mathrm{KLD}_{\mathrm{r},t}^{\mathcal{H}_0\rightarrow \mathcal{H}_1}=\mathrm{KLD}_{\mathrm{r},t}^{\mathcal{H}_1\rightarrow \mathcal{H}_0}$. The average KLD of the whole ISAC system can be formulated as follows,
\begin{equation}
    \mathrm{KLD}_{\mathrm{ISAC}}= \frac{1}{K}\sum_{k=1}^{K}\mathrm{KLD}_{\mathrm{c},k}+\frac{1}{T}\sum_{t=1}^{T}\mathrm{KLD}_{\mathrm{r},t}.
\end{equation}
This formulation represents the aggregate information-theoretic performance measure across both subsystems, where the averaging within each subsystem ensures balanced treatment regardless of the number of users $K$ or targets $T$, where $\mathrm{KLD}_{\mathrm{ISAC}}$ provides a unified optimisation objective that captures the fundamental goal of maximising information extraction from shared ISAC resources.

\vspace{-0.1 in}
\section{Radar Waveform Optimisation \label{sec:radop}}
 The KLD-based optimisation framework provides analytically tractable KLD expressions, and their gradients as shown in  Sec. \ref{sec:CommKLD}, Sec. \ref{sec:kldrad}, and Appendix A. This computational efficiency, combined with the unified performance characterisation, makes KLD particularly suitable for real-time ISAC applications. In this section, the radar waveform is optimised while using the ZF beamforming scheme for the communication subsystem. The optimisation problem is,
\vspace{-0.05 in}
\begin{subequations}
\label{eq:op2}
\begin{align}
\mathcal{P}_{1}:\max_{\mathbf{\dot{W}}_{\mathrm{r}}} \quad\!\! & \frac{1}{T}\sum_{t=1}^{T}\mathrm{KLD}_{\mathrm{r},t|\mathbf{W}_{\mathrm{r},t},\mathbf{H}_t} \label{10.a}\\
\textrm{s.t.}\quad &  {\mathrm{{KLD}}_{\mathrm{r},t}} \geq \underline{A}_t,  \,\,\,  \forall t \in T,  \label{10.b}\\
  &{\mathrm{{KLD}}_{\mathrm{c},k}} \geq \underline{B}_k,      \forall k \in K,   \label{10.c}  \\
  & \sum_{t=1}^{T}\frac{1}{L}\|\mathbf{W}_{\mathrm{r},t}\|_F^2 \leq P_\mathrm{r}, \label{10.d} \\
  & P_\mathrm{c} =P_\mathrm{T} -P_\mathrm{r}, \label{10.e}
\end{align}
\end{subequations}
where $\mathrm{KLD}_{\mathrm{r},t|\mathbf{W}_{\mathrm{r},t},\mathbf{H}_t}$ is defined in \eqref{kldrt}. After substituting \eqref{kldrt} in \eqref{10.a}, can be written as $\mathcal{P}_{2}$ that is given on page \pageref{rrr}. This optimisation problem focuses on the radar waveform while considering the whole ISAC system performance. The minimum KLD requirements in constraints \eqref{10.b} and \eqref{10.c} are carefully chosen to ensure specific performance guarantees in terms of detection probability $P_D$ and BER for the radar and communication subsystems, respectively. For example, for the radar subsystem, a KLD threshold of $\underline{A}_t = 10$ bits corresponds to a detection probability of approximately $P_D \approx 0.25$, representing a baseline detection capability. For the communication subsystem, a KLD threshold of $\underline{B}_k = 10$ bits ensures a BER of approximately $10^{-2}$ for QPSK modulation, which is acceptable for many practical applications. These thresholds can be adjusted based on specific system requirements—higher KLD values would yield improved performance at the cost of more stringent resource allocation. The mathematical relationships between KLD and these conventional metrics, established in \cite{Al-Jarrah2023}, allow system designers to set appropriate thresholds based on application needs. The constraint on radar power \eqref{10.d} is explicit, where fixed power allocation is used between subsystems, because the radar waveform is our primary optimisation variable. The communication power is implicitly considered through the ZF beamforming scheme and the constraint \eqref{10.c} on communication performance. Specifically, the communication subsystem utilises its allocated power through the ZF beamforming scheme to distribute power to each UE, with weights predetermined based on the channel state information.
\begin{table*}
\vspace{-0.1in}
\begin{minipage}{1\textwidth}
\vspace{-0.1in}
\begin{equation}
\begin{aligned}
\mathcal{P}_{2}: \max_{\mathbf{\dot{W}}_{\mathrm{r}}} \quad &\!\!\!\! \frac{1}{T}\sum_{t=1}^{T}\frac{1}{\ln{2}}\Bigg(\!\ln\!\left(\mathrm{det}(\:d_{\mathrm{r},t}^{\zeta}\mathbf{H}_t(\frac{1}{L}\mathbf{W}_{\mathrm{r},t}\mathbf{W}_{\mathrm{r},t}^H+\mathbf{W}_{\mathrm{c}}\mathbf{W}_{\mathrm{c}}^H)\mathbf{H}_t^H+\sigma_n^2\mathbf{I}_N)\!\right)\\
&+\mathrm{Tr}\left(\left(\:d_{\mathrm{r},t}^{\zeta}\mathbf{H}_t(\frac{1}{L}\mathbf{W}_{\mathrm{r},t}\mathbf{W}_{\mathrm{r},t}^H+\mathbf{W}_{\mathrm{c}}\mathbf{W}_{\mathrm{c}}^H)\mathbf{H}_t^H+\sigma_n^2\mathbf{I}_N\right)^{-1}\!\sigma_n^2\!\right)\! -\!N\!\left(1+\ln\left(\sigma_n^2\right)\right)\!\Bigg)\\
\textrm{s.t.} \quad\quad\quad\quad & \text{\eqref{10.b}, \eqref{10.c}, \eqref{10.d}, and \eqref{10.e}}
  \label{eq:op3}\\
\end{aligned}
\end{equation}
\vspace{-0.14in}
\medskip
\hrule
\vspace{-0.19in}
\end{minipage}\label{rrr}
\end{table*}
To solve the optimisation problem in \eqref{eq:op3}, we utilise the Projected Gradient method with a penalty function \cite{nocedal2006numerical}. This method is chosen for its suitability to the non-linear, non-convex optimisation problem \cite{bertsekas1999nonlinear}. This method uses a penalty function for KLD constraints and handles radar power via projection, balancing the solution quality and computational efficiency for real-time ISAC applications. Its scalability makes it suitable for radar waveform optimisation in ISAC. The optimisation variable $\mathbf{\dot{W}}_\mathrm{r} \in \mathbb{C}^{N \times L \times T}$ is a three-dimensional tensor, where $\mathbf{w}_{\mathrm{r},t,l}$, $\mathbf{W}_{\mathrm{r},t}$, and $\mathbf{W}_{\mathrm{r},l}$ are all parts of this tensor. We define the objective function $f(\mathbf{\dot{W}}_\mathrm{r})$ and a penalty function $p(\mathbf{\dot{W}}_\mathrm{r})$ to handle the constraints in \eqref{10.b} and \eqref{10.c},
\vspace{-0.05 in}
\begin{equation}
f(\mathbf{\dot{W}}_\mathrm{r}) = \frac{1}{T}\sum_{t=1}^T \mathrm{KLD}_{\mathrm{r},t|\mathbf{W}_{\mathrm{r},t},\mathbf{H}_t},
\label{objfunrad}
\end{equation}
\vspace{-0.1 in}
\begin{equation}
\begin{split}
p(\mathbf{\dot{W}}_\mathrm{r}) = & \sum_{t=1}^T \left(\max(0, \underline{A}_t - \mathrm{KLD}_{\mathrm{r},t|\mathbf{W}_{\mathrm{r},t},\mathbf{H}_t})^2 \right) \\
& + \sum_{k=1}^K \left(\max(0, \underline{B}_k - \mathrm{KLD}_{\mathrm{c},k|\mathbf{\dot{W}}_{\mathrm{r}}})^2 \right).
\end{split}
\end{equation}
The gradient of the objective function, $\nabla f(\mathbf{\dot{W}}_\mathrm{r}) \in \mathbb{C}^{N \times L \times T}$, is a gradient tensor computed element-wise, as detailed in Appendix~A The final equation is given by,
\vspace{-0.05 in}
\begin{equation}
\begin{split}
[\nabla f(\mathbf{\dot{W}}_\mathrm{r})]_{i,l,t} = & \frac{2\:d_{\mathrm{r},t}^{\zeta}}{TL\ln{2}}\Big(\mathbf{H}_t^H(\mathbf{R}_{2,t})^{-1}\mathbf{H}_t\mathbf{W}_{\mathrm{r},t} \\
& -\sigma_n^2(\mathbf{R}_{2,t})^{-2}\mathbf{H}_t\mathbf{W}_{\mathrm{r},t}\Big)_i. \label{41}
\end{split}
\end{equation}
The gradient of the penalty function $\nabla p(\mathbf{\dot{W}}_\mathrm{r}) \in \mathbb{C}^{N \times L \times T}$ is also a gradient tensor, computed as,
\vspace{-0.05 in}
\begin{equation}
\begin{split}
[\nabla p(\mathbf{\dot{W}}_\mathrm{r})]_{i,l,t} = & -2(\underline{A}_t - \mathrm{KLD}_{\mathrm{r},t})[\nabla \mathrm{KLD}_{\mathrm{r},t}]_{i,l} \cdot \mathbb{I}_1 \\
& - \sum_{k=1}^K 2(\underline{B}_k - \mathrm{KLD}_{\mathrm{c},k})[\nabla \mathrm{KLD}_{\mathrm{c},k}]_{i,l,t} \cdot \mathbb{I}_2, \label{42}
\end{split}
\end{equation}
where $\mathbb{I}_1 = \mathbb{I}(\mathrm{KLD}_{\mathrm{r},t} < \underline{A}_t)$ and $\mathbb{I}_2 = \mathbb{I}(\mathrm{KLD}_{\mathrm{c},k} < \underline{B}_k)$ with $\mathbb{I}(\cdot)$ represents the indicator function. The gradients of $\mathrm{KLD_\mathrm{r}}$ and $\mathrm{KLD_\mathrm{c}}$ with respect to  $\mathbf{W}_{\mathrm{r},t}$ and $\mathbf{\dot{W}}_\mathrm{r}$, respectively, are provided in Appendix~A. The final forms are given by,
\vspace{-0.05 in}
\begin{align}
[\nabla \mathrm{KLD}_{\mathrm{r},t|\mathbf{W}_{\mathrm{r},t},\mathbf{H}_t}]_{i,l} = & \frac{2\:d_{\mathrm{r},t}^{\zeta}}{L\ln{2}}\Big(\mathbf{H}_t^H\mathbf{R}_{2,t}^{-1}\mathbf{H}_t\mathbf{W}_{\mathrm{r},t} \nonumber \label{43}\\
& - \sigma_n^2\mathbf{H}_t^H\mathbf{R}_{2,t}^{-2}\mathbf{H}_t\mathbf{W}_{\mathrm{r},t}\Big)_i , \\
[\nabla \mathrm{KLD}_{\mathrm{c},k|\mathbf{\dot{W}}_{\mathrm{r}},\mathbf{H}_t}]_{i,l,t} \!=\! & -\!\frac{2d_{\mathrm{c},k}^{\zeta}\lambda p_k (N\!\!-\!\!K)\sigma_h^2}{M(M\!\!-\!\!1) L \sigma_{\eta}^2\ln 2}[\mathbf{\dot{W}}_{\!\mathrm{r}}]_{i,l,t} \label{44}.
\end{align}
The update rule for the projected gradient method with penalty ensures that each iteration moves the solution towards optimality while maintaining feasibility, and can be shown as follows,
\vspace{-0.05 in}
\begin{equation}
\mathbf{\dot{W}}_\mathrm{r}^{(n+1)} = \mathcal{P}\left(\mathbf{\dot{W}}_\mathrm{r}^{(n)} + \alpha_n\mathbf{G}^{(n)}\right),
\end{equation}
where $\mathbf{G}^{(n)} = \nabla f(\mathbf{\dot{W}}_\mathrm{r}^{(n)}) - \rho_n \nabla p(\mathbf{\dot{W}}_\mathrm{r}^{(n)})$ is the gradient direction, combining the objective function gradient and the penalty function gradient, $\rho_n$ is the penalty parameter, balancing the objective and constraint satisfaction, $\alpha_n$ is the step size, and $\mathcal{P}$ is the projection onto the feasible set defined by the power constraint, ensuring that the power constraint in \eqref{10.d} is always satisfied. The projection mechanism is shown as,
\vspace{-0.03 in}
\begin{equation}
\mathcal{P}(\mathbf{\dot{W}}_\mathrm{r}) = \begin{cases}
    \mathbf{\dot{W}}_\mathrm{r} \sqrt{\frac{P_\mathrm{r}L}{\|\mathbf{\dot{W}}_\mathrm{r}\|_F^2}}, & \text{if } \frac{1}{L}\|\mathbf{\dot{W}}_\mathrm{r}\|_F^2 > P_\mathrm{r} \\
    \mathbf{\dot{W}}_\mathrm{r}, & \text{otherwise} \label{46}
\end{cases}
\end{equation}
\begin{figure}[t]
\vspace{-0.2in}
\begin{algorithm}[H]
\caption{Projected Gradient Method with Penalty for Radar Waveform Optimisation}
\label{alg1}
\begin{algorithmic}[1]
\REQUIRE{Initial point $\mathbf{\dot{W}}_\mathrm{r}^{(0)}$, constants $\alpha_0$, $\rho_0$, $\beta$, $\gamma$, $c$, $P_\mathrm{r}$, $\underline{A}_t$, $\underline{B}_k$, tolerance $\varepsilon$, and maximum iterations $\mathrm{max\_iter}$}
\ENSURE{Optimal solution $\mathbf{\dot{W}}_\mathrm{r}^*$}
\STATE{Initialise $n = 0$}
\WHILE{$n < \mathrm{max\_iter}$}
\STATE{Compute gradients $\nabla f(\mathbf{\dot{W}}_\mathrm{r}^{(n)})$ and $\nabla p(\mathbf{\dot{W}}_\mathrm{r}^{(n)})$}
\STATE{Compute total gradient $\mathbf{G}^{(n)} = \nabla f(\mathbf{\dot{W}}_\mathrm{r}^{(n)}) - \rho_n \nabla p(\mathbf{\dot{W}}_\mathrm{r}^{(n)})$}
\STATE{Perform backtracking line search to find $\alpha_n$}
\STATE{Update $\mathbf{\dot{W}}_\mathrm{r}^{(n+1)} = \mathcal{P}(\mathbf{\dot{W}}_\mathrm{r}^{(n)} + \alpha_n\mathbf{G}^{(n)})$}
\STATE{Compute $f(\mathbf{\dot{W}}_\mathrm{r}^{(n+1)})$ and $p(\mathbf{\dot{W}}_\mathrm{r}^{(n+1)})$}
\STATE{Update $\rho_{n+1}$ according to the penalty update rule}
\IF{$\|\mathbf{\dot{W}}_\mathrm{r}^{(n+1)} - \mathbf{\dot{W}}_\mathrm{r}^{(n)}\|_F < \varepsilon$}
\STATE{\textbf{break}}
\ENDIF
\STATE{$n = n + 1$}
\ENDWHILE
\RETURN{$\mathbf{\dot{W}}_\mathrm{r}^* = \mathbf{\dot{W}}_\mathrm{r}^{(n)}$}
\end{algorithmic}
\end{algorithm}
\vspace{-0.36in}
\end{figure}
The step size $\alpha_n$ is determined using a backtracking line search to ensure convergence. We find the smallest non-negative integer $m$ such that,
\vspace{-0.03 in}
\begin{equation}
\begin{split}
& f(\mathbf{\dot{W}}_\mathrm{r}^{(n+1)}) - \rho_n p(\mathbf{\dot{W}}_\mathrm{r}^{(n+1)}) \geq \\
& f(\mathbf{\dot{W}}_\mathrm{r}^{(n)}) - \rho_n p(\mathbf{\dot{W}}_\mathrm{r}^{(n)}) + c\alpha_n\|\mathbf{G}^{(n)}\|_F^2,
\end{split}
\end{equation}
where $c \in (0,1)$. This condition, known as the Armijo condition, ensures a sufficient decrease in the penalised objective function \cite{armijo1966}. The step size is set as $\alpha_n = \beta^m \alpha_{n-1}$, with $\beta \in (0,1)$ as a reduction factor. The penalty parameter $\rho_n$ is updated as follows,
\vspace{-0.05 in}
\begin{equation}
\rho_{n+1} = \begin{cases}
    \gamma \rho_n, & \text{if } p(\mathbf{\dot{W}}_\mathrm{r}^{(n+1)}) > 0 \\
    \rho_n, & \text{otherwise} \label{48}
\end{cases}
\end{equation}
where $\gamma > 1$ is a constant factor. This adaptive scheme increases the penalty when constraints are violated and keeps it constant when they are satisfied. The projection operator $\mathcal{P}$ ensures that the power constraint is always satisfied. The backtracking line search procedure is used to determine an appropriate step size $\alpha_n$ at each iteration. This adaptive step size selection helps to ensure a sufficient increase in the objective function while maintaining the stability of the algorithm. The line search condition can be expressed as,
\vspace{-0.05 in}
\begin{equation}
\begin{split}
\!\!\phi(\mathbf{\dot{W}}_\mathrm{\!r}^{(n+1)}) \!\!\leq & \phi(\mathbf{\dot{W}}_\mathrm{\!r}^{\!(n)}) \! +\! c\alpha_n\!\langle \nabla\! \phi(\mathbf{\dot{W}}_\mathrm{r}^{\!(n)}),\! \mathbf{\dot{W}}_\mathrm{r}^{(n+1)} \!\!-\!\! \mathbf{\dot{W}}_\mathrm{r}^{(n)} \rangle,\!\!\!
\end{split}
\end{equation}
where $\phi(\mathbf{\dot{W}}_\mathrm{\!r}) \!= \!\!f(\mathbf{\dot{W}}_\mathrm{\!r}) \!-\!\! \rho_{n} p(\mathbf{\dot{W}}_\mathrm{\!r})$ is the penalised objective function. The penalty parameter $\rho_n$ is updated adaptively based on the constraint violation, allowing the algorithm to balance optimisation and constraint satisfaction. If constraints are consistently violated, the increasing penalty will emphasise constraint adherence.
\vspace{-0.1 in}

\subsection{Radar waveform detection probability optimisation \label{D-ROP}}

To incorporate the detection performance directly into the optimisation framework, we derive the probability of detection $P_{D,t}$ from the GLRT in Sec. \ref{RAD-ANA}.A, where $\mathbf{Q}_t = \mathbf{I}_N + \mathbf{R}_{1,t}^{-1}$. For large $L$ and by invoking the CLT, the test statistic converges to a Gaussian distribution under both hypotheses. Under $\mathcal{H}_0$ where $\mathbf{y}_{\mathrm{r},t,l} = \mathbf{n}_{t,l}$, the mean and variance are $\mu_0 = \sigma_n^2\text{tr}(\mathbf{Q}_t)$ and $\sigma_0^2 = \sigma_n^4\text{tr}(\mathbf{Q}_t^2)/L$. Under $\mathcal{H}_1$ where $\mathbf{y}_{\mathrm{r},t,l} = d_{\mathrm{r},t}^{\zeta/2}\mathbf{H}_t\mathbf{x}_{t,l} + \mathbf{n}_{t,l}$, we have $\mu_1 = \mu_0 + \Delta\mu$ and $\sigma_1^2 = \sigma_0^2 + 2\sigma_n^2\Delta\mu/L$, where $\Delta\mu = d_{\mathrm{r},t}^{\zeta}\text{tr}(\mathbf{H}_t\mathbf{R}_t\mathbf{H}_t^H\mathbf{Q}_t)$. For a fixed false alarm rate $P_{FA}$, the Neyman-Pearson threshold is,
\vspace{-0.05 in}
\begin{equation}
\tau_2 = \mu_0 + Q^{-1}(P_{FA})\sigma_0.
\label{eq:np_threshold}
\end{equation}
Substituting \eqref{eq:np_threshold} into the detection probability expression $P_{D,t} = Q((\tau_2 - \mu_1)/\sigma_1)$ yields \cite{Al-Jarrah2023},
\vspace{-0.05 in}
\begin{equation}
P_{D,t|\mathbf{W}_{\mathrm{r},t},\mathbf{H}_t} = Q\left(\frac{Q^{-1}(P_{FA})\sigma_0 - \Delta\mu}{\sigma_1}\right),
\label{eq:pd_final}
\end{equation}
where $Q(\cdot)$ is the complementary cumulative distribution function of the standard normal distribution. The detection probability in \eqref{eq:pd_final} depends on the waveform design through $\mathbf{R}_t$, which affects both $\mathbf{Q}_t$ and $\Delta\mu$. We form a baseline radar waveform optimisation by using the detection probability directly in the problem defined in \eqref{eq:op12} below. 
\vspace{-0.09 in}
\begin{subequations}
\label{eq:op12}
\begin{align}
\mathcal{P}_{3}:\max_{\mathbf{\dot{W}}_{\mathrm{r}}} \quad\!\! & \frac{1}{T}\sum_{t=1}^{T}P_{D,t|\mathbf{W}_{\mathrm{r},t},\mathbf{H}_t}  \label{12.a}\\
\textrm{s.t.}\quad &  \eqref{10.c}, \eqref{10.d}, \eqref{10.e} \nonumber\\
  &P_{D,t|\mathbf{W}_{\mathrm{r},t},\mathbf{H}_t}  \leq \overline{D}_t\,      \forall t \in T,   \label{12.c}
\end{align}
\end{subequations}
where the communication subsystem uses ZF beamforming with precoding matrix $\mathbf{W}_\mathrm{c}$  and $\overline{D}_{t}$ is the upper limit of the $P_D$ of the $t$-th target. This $P_{D,t}$ objective depends on the $Q(\cdot)$ function and its inverse, which introduces additional computational complexity in the optimisation process due to its non-elementary nature and the challenges in evaluating its derivatives efficiently. This distinction illustrates the enhanced analytical tractability of the proposed KLD-based approach. This problem is inherently non-convex and nonlinear due to the $Q$-function and its constraint. Using $\nabla_{\mathbf{W}_{\mathrm{r},t}}\,\mathrm{P}_{D,t}$ derived in Appendix~B, and the gradients derived previously, we employ the same projected gradient method with penalty as in Algorithm~\ref{alg1}, to handle the nonconvexities in \eqref{eq:op12}.
\vspace{-0.09 in}
\section{Communication optimisation \label{com-op}}

The communication subsystem optimisation uses the gradient-assisted IPM \cite{doi:10.1137/S0036144502414942}, in contrast to the projected gradient method with penalty used for radar optimisation, due to the different optimisation variable structures. Radar optimisation involves a three-dimensional tensor $\mathbf{\dot{W}}_\mathrm{r} \in \mathbb{C}^{N \times L \times T}$, while communication beamforming optimises a two-dimensional matrix $\mathbf{W}_\mathrm{c} \in \mathbb{C}^{N \times K}$. This reduced dimensionality makes IPM particularly suitable, potentially offering faster convergence and more precise solutions. The communication waveform is optimised while using the conventional identity covariance (CIC) matrix design for the radar subsystem, with $\mathbf{R}_{w}=\mathbf{I}_{N\times N} \quad \forall t \in T$. The optimisation problem can be expressed as follows,
\vspace{-0.05 in}
\begin{subequations}
\begin{align}
\mathcal{P}_{4}:\max_{\mathbf{W}_{\mathrm{c}}} \quad\!\! & \frac{1}{K}\sum_{k=1}^{K}\mathrm{KLD}_{\mathrm{c},k|\mathbf{W}_{\mathrm{c}},\mathbf{h}_k}\label{11.a}\\
\textrm{s.t.}\quad &  \eqref{10.b}, \eqref{10.c} \nonumber\\
  & \|\mathbf{W}_{\mathrm{c}}\|_F^2 \leq P_\mathrm{c}, \label{11.d} \\
  & P_\mathrm{r} =P_\mathrm{T} -P_\mathrm{c}. \label{11.e}\vspace{-0.05 in}
\end{align}
\end{subequations}
 Similar to the radar waveform optimisation, fixed power allocation is used between subsystems, where radar power is uniformly distributed among potential targets, as the radar covariance matrix ensures omnidirectional beam flow. To solve this optimisation problem, we must derive the gradients of the objective function and the constraints. Let’s define the objective function as follows,
\vspace{-0.05 in}
\begin{equation}
f(\mathbf{W}_{\mathrm{c}}) = \frac{1}{K} \sum_{k=1}^{K} \mathrm{KLD}_{\mathrm{c},k|\mathbf{W}_{\mathrm{c}},\mathbf{h}_k}.
\end{equation}
To compute the gradient of the objective function, we need to derive the gradients of each component of $\mathbf{W}_\mathrm{c}$. Let $a_k = d_{\mathrm{c},k}^{\zeta}|\mathbf{h}_k^T \mathbf{w}_{\mathrm{c},k}|^2$. Using the chain rule and properties of matrix derivatives, the gradients can be expressed as follows,
\vspace{-0.05 in}
\begin{align}
\nabla_{\mathbf{w}_{\mathrm{c},j}} |\mathbf{h}_k^T \mathbf{w}_{\mathrm{c},j}|^2 &= 2d_{\mathrm{c},k}^{\zeta} \mathbf{h}_k \mathbf{h}_k^T \mathbf{w}_{\mathrm{c},j}, \quad\quad\quad\quad \forall j
\label{52}\\
\nabla{\mathbf{w}_{\mathrm{c},j}} \mathrm{KLD}_{\mathrm{c},k} &=
\begin{cases}
C \cdot \frac{2d_{\mathrm{c},k}^{\zeta} \mathbf{h}_k \mathbf{h}_k^T \mathbf{w}_{\mathrm{c},k}}{b_k}, & \text{if } j = k \\
-C \cdot \frac{2d_{\mathrm{c},k}^{\zeta} \mathbf{h}_k \mathbf{h}_k^T \mathbf{w}_{\mathrm{c},j} \cdot a_k}{b_k^2}, & \text{if } j \neq k
\end{cases}
\end{align}
where $C = \frac{\lambda}{2 M(M-1)\ln 2}$, and $b_k = d_{\mathrm{c},k}^{\zeta}\sum_{i=1, i \neq k}^{K} \|\mathbf{h}_k^T \mathbf{w}_{\mathrm{c},i}\|^2 + \sigma_{\eta}^2$. Finally, we can express the gradient of the objective function as,
\vspace{-0.05 in}
\begin{equation}
\begin{split}
\nabla_{\mathbf{w}_{\mathrm{c},k}} f(\mathbf{W}_{\mathrm{c}}) = \frac{ d_{\mathrm{c},k}^{\zeta}C}{K} \sum_{k=1}^{K} \Bigg(\frac{2 \mathbf{h}_k \mathbf{h}_k^T \mathbf{w}_{\mathrm{c},k}}{b_k} \\
- \sum_{j \neq k} \frac{2 \mathbf{h}_k \mathbf{h}_k^T \mathbf{w}_{\mathrm{c},j} \cdot a_k}{b_k^2} \Bigg).
\end{split}
\end{equation}
Next, we need to consider the constraints of our optimisation problem. The power constraint is given by,
\vspace{-0.05 in}
\begin{equation}
g(\mathbf{W}_{\mathrm{c}}) = \|\mathbf{W}_{\mathrm{c}}\|_F^2 - P_{\mathrm{c}} \leq 0.\label{55}\vspace{-0.05 in}
\end{equation}
The gradient of this constraint is derived as follows,
\vspace{-0.05 in}
\begin{equation}
\nabla_{\mathbf{w}_{\mathrm{c},j}} g(\mathbf{W}_{\mathrm{c}}) = 2 \mathbf{w}_{\mathrm{c},j}.\vspace{-0.05 in}
\end{equation}
The gradient of the radar KLD constraint with respect to $\mathbf{W}_{\mathrm{c}}$ is derived as follows,
\vspace{-0.05 in}
\begin{equation}
\nabla_{\mathbf{W}_{\mathrm{c}}} \mathrm{KLD}_{\mathrm{r},t} = \frac{d_{\mathrm{r},t}^{\zeta}}{\ln 2} \left( \mathbf{R}_{2,t}^{-1} \mathbf{H}_t \mathbf{W}_{\mathrm{c}} \mathbf{H}_t^H \right). \label{57}\vspace{-0.03 in}
\end{equation}
\begin{figure}[t]
\vspace{-0.12 in}
\begin{algorithm}[H]
\caption{Interior Point Method for Communication Optimisation}
\label{alg2}
\begin{algorithmic}[1]
\REQUIRE{Initial point $\mathbf{W}_\mathrm{c}^{(0)}$, constants $\mu_0$, $\gamma$, $P_\mathrm{c}$, $\underline{A}_t$, $\underline{B}_k$, tolerance $\epsilon$, and maximum iterations $\mathrm{max\_iter}$}
\ENSURE{Optimal solution $\mathbf{W}_\mathrm{c}^*$}
\STATE{Initialise $n = 0$, $\mu = \mu_0$}
\WHILE{$n < \mathrm{max\_iter}$}
\STATE{Compute $f(\mathbf{W}_\mathrm{c}^{(n)})$, $\nabla f(\mathbf{W}_\mathrm{c}^{(n)})$, $g_i(\mathbf{W}_\mathrm{c}^{(n)})$, and $\nabla g_i(\mathbf{W}_\mathrm{c}^{(n)})$}
\STATE{Compute KKT conditions}
\IF{KKT conditions satisfied within $\epsilon$}
\STATE{\textbf{break}}
\ENDIF
\STATE{Solve Newton system for search direction $\Delta \mathbf{W}_\mathrm{c}$}
\STATE{Perform line search to find step size $\alpha$}
\STATE{Update $\mathbf{W}_\mathrm{c}^{(n+1)} = \mathbf{W}_\mathrm{c}^{(n)} + \alpha \Delta \mathbf{W}_\mathrm{c}$}
\STATE{Update $\mu = \gamma \mu$}
\STATE{$n = n + 1$}
\ENDWHILE
\RETURN{$\mathbf{W}_\mathrm{c}^* = \mathbf{W}_\mathrm{c}^{(n)}$}
\end{algorithmic}
\end{algorithm}
\vspace{-0.34in}
\end{figure}
Solving the optimisation problem defined in equations \eqref{11.a}-\eqref{11.d} uses the IPM, as detailed in \textbf{Algorithm \ref{alg2}}. This method efficiently addresses constrained optimisation with key parameters: an initial barrier parameter \(\tilde{\mu_0}\), a barrier reduction factor \(\tilde{\gamma}\), and a convergence tolerance \(\epsilon\). The algorithm iteratively computes the objective function, constraints, and their gradients, checks the Karush-Kuhn-Tucker (KKT) conditions and updates the solution using a Newton system and line search. The barrier parameter \(\tilde{\mu}\) is updated as \(\tilde{\mu} = \tilde{\gamma}\tilde{\mu}\) in each iteration, gradually enforcing constraints more strictly. This continues until \( \|\mathbf{W}_\mathrm{c}^{(n+1)} - \mathbf{W}_\mathrm{c}^{(n)}\|_F < \epsilon \) or the maximum iterations are reached, yielding the optimal beamforming matrix \(\mathbf{W}_\mathrm{c}^*\) that maximises the communication KLD while satisfying radar and power constraints.
\vspace{-0.1in}
\subsection{Communication waveform BER-optimisation \label{B-COP}}
\label{subsec:baseline_ber}
We formulate another baseline design that minimises the instantaneous BER with respect to $\mathbf{W}_{\mathrm{c}}$, while ensuring a minimum radar KLD per target. Where the instantaneous BER of user~$k$ is approximated by \eqref{MPSK_BER}. The optimisation problem is,
\vspace{-0.05 in}
\begin{subequations}\label{BERlimitOPT}
\begin{align}
\mathcal{P}_{5}:\min_{\mathbf{W}_{\mathrm{c}}} \quad\!\! & \frac{1}{K}\sum_{k=1}^{K}\mathrm{BER}_{k|\mathbf{W}_{\mathrm{c}},\mathbf{h}_k}\label{BERlimit-obj}\\
\textrm{s.t.}\quad &  \eqref{10.b}, \eqref{11.d}, \eqref{11.e} \nonumber\\
  &\mathrm{BER}_{k}\;\le\;\overline{C}_k,      \forall k \in K,   
\end{align}
\end{subequations}
where the radar subsystem precoder $\mathbf{W}_\mathrm{r}$ uses the identity covariance $\mathbf{R}_w = \mathbf{I}_N$. As such, it is important to define the BER, where for $M$-ary constellations, BER of user~$k$ is approximated by \cite{proakis2008digital},
\vspace{-0.05 in}
\begin{equation}
\mathrm{BER}_{k}
~\approx~
\frac{\lambda_\mathrm{c}}{M \,\log_2 M} \,
Q\!\Bigl(\sqrt{\beta_\mathrm{c}\,\mathrm{SINR}_{k}}\Bigr),
\label{MPSK_BER}
\end{equation}
where $\lambda_\mathrm{c}$ and $\beta_\mathrm{c}$ are constellation-dependent constants (e.g., $\lambda_\mathrm{c}=4$, $\beta_\mathrm{c}=2$ for QPSK). Unlike the unified closed-form gradients in our KLD-based design, this BER objective depends on the $Q(\cdot)$ function in \eqref{MPSK_BER}, which introduces additional computational complexity in the optimisation process due to its non-elementary nature and the challenges in evaluating its derivatives efficiently. This distinction illustrates the enhanced analytical tractability of the proposed KLD-based approach. This problem is inherently non-convex and nonlinear due to the $Q$-function in \eqref{MPSK_BER} and $\mathrm{KLD}_{\mathrm{r},t}$ constraint. Using $\nabla_{\mathbf{W}_{\mathrm{c}}}\,\mathrm{BER}_{k}$ derived in Appendix~C, and the constraint gradients derived in \eqref{55}-\eqref{57}, we employ the same gradient-assisted IPM as in Algorithm~\ref{alg2}, to handle the nonconvexities in \eqref{BERlimitOPT}, leveraging the reduced dimensionality of $\mathbf{W}_{\mathrm{c}}\in\mathbb{C}^{N\times K}$. 
\vspace{-0.1in}

\section{Integrated Waveform Optimisation for ISAC \label{int-op}}
Finally, we introduce this unified approach that maximises the whole ISAC system KLD, which ensures improved radar detection and communication reliability while satisfying minimum subsystem KLD constraints and adhering to a predefined power budget. The optimisation problem is expressed as,
\begin{subequations}
\label{eq:unified_opt}
\begin{align}
\mathcal{P}_{6}: \max_{\mathbf{W}_{\mathrm{r}},\mathbf{W}_{\mathrm{c}}} \quad & \mathrm{KLD}_{\mathrm{ISAC}}(\mathbf{W}_{\mathrm{r}},\mathbf{W}_{\mathrm{c}}) \label{eq:unified_obj} \\
\text{s.t.} \quad & \eqref{10.b}, \eqref{10.c} \nonumber\\
& \|\mathbf{W}_{\mathrm{c}}\|_F^2 + \sum_{t=1}^{T} \frac{1}{L} \|\mathbf{W}_{\mathrm{r},t}\|_F^2 \leq P_\mathrm{T}. \label{eq:unified_const3} \\
  & P_\mathrm{T} =P_\mathrm{c} +P_\mathrm{r}.
\end{align}
\end{subequations}
The total power constraint in \eqref{eq:unified_const3} limits the combined radar and communication power to $P_\mathrm{T}$, and optimises the power allocation between the two subsystems. To solve this non-convex problem, the ADMM is employed. This approach unifies the projected gradient method used for radar waveform optimisation and the IPM technique used for communication waveform optimisation, efficiently handling the coupling between subsystems while decomposing the complex problem into more manageable subproblems. This involves introducing auxiliary variables $\mathbf{Z}_{\mathrm{r}}$ and $\mathbf{Z}_{\mathrm{c}}$ such that $\mathbf{Z}_{\mathrm{r}} = \mathbf{W}_{\mathrm{r}}$ and $\mathbf{Z}_{\mathrm{c}} = \mathbf{W}_{\mathrm{c}}$. The augmented Lagrangian for the problem is given as,
\begin{align}
\mathcal{L}_{\rho} = & \frac{1}{T} \sum_{t=1}^{T} \mathrm{KLD}_{\mathrm{r},t} + \frac{1}{K} \sum_{k=1}^{K} \mathrm{KLD}_{\mathrm{c},k} \nonumber \\
& + \langle \mathbf{U}_{\mathrm{r}}, \mathbf{W}_{\mathrm{r}} - \mathbf{Z}_{\mathrm{r}} \rangle + \langle \mathbf{U}_{\mathrm{c}}, \mathbf{W}_{\mathrm{c}} - \mathbf{Z}_{\mathrm{c}} \rangle \nonumber \\
& - \frac{\rho}{2} \left( \|\mathbf{W}_{\mathrm{r}} - \mathbf{Z}_{\mathrm{r}}\|_F^2 + \|\mathbf{W}_{\mathrm{c}} - \mathbf{Z}_{\mathrm{c}}\|_F^2 \right),\vspace{-0.05 in}
\end{align}
where $\mathbf{U}_{\mathrm{r}}$ and $\mathbf{U}_{\mathrm{c}}$ are the dual variables (Lagrange multipliers), and $\rho > 0$ is the ADMM penalty parameter. The ADMM algorithm proceeds iteratively, updating the primal variables $(\mathbf{W}_{\mathrm{r}}, \mathbf{W}_{\mathrm{c}})$, auxiliary variables $(\mathbf{Z}_{\mathrm{r}}, \mathbf{Z}_{\mathrm{c}})$, and dual variables $(\mathbf{U}_{\mathrm{r}}, \mathbf{U}_{\mathrm{c}})$. The updates are given as follows,
\begin{align}
\mathbf{W}_{\mathrm{r}}^{(n+1)} &= \mathbf{Z}_{\mathrm{r}}^{(n)} - \frac{1}{\rho} \mathbf{U}_{\mathrm{r}}^{(n)}, \label{up1}\\ 
\mathbf{W}_{\mathrm{c}}^{(n+1)} &= \mathbf{Z}_{\mathrm{c}}^{(n)} - \frac{1}{\rho} \mathbf{U}_{\mathrm{c}}^{(n)}, \label{up2}\\
\mathbf{Z}_{\mathrm{r}}^{(n+1)} &= \arg \min_{\mathbf{Z}_{\mathrm{r}}} \mathcal{L}_{\rho}, \quad \label{up3}
\mathbf{Z}_{\mathrm{c}}^{(n+1)} = \arg \min_{\mathbf{Z}_{\mathrm{c}}} \mathcal{L}_{\rho}, \\
\mathbf{U}_{\mathrm{r}}^{(n+1)} &= \mathbf{U}_{\mathrm{r}}^{(n)} + \rho (\mathbf{W}_{\mathrm{r}}^{(n+1)} - \mathbf{Z}_{\mathrm{r}}^{(n+1)}), \label{up4}\\
\mathbf{U}_{\mathrm{c}}^{(n+1)} &= \mathbf{U}_{\mathrm{c}}^{(n)} + \rho (\mathbf{W}_{\mathrm{c}}^{(n+1)} - \mathbf{Z}_{\mathrm{c}}^{(n+1)}). \label{up5}\vspace{-0.05 in}
\end{align}
The auxiliary variables are updated by solving the following constrained subproblem,
\begin{align}
\vspace{-0.05 in}
\min_{\mathbf{Z}_{\mathrm{r}}, \mathbf{Z}_{\mathrm{c}}} \quad & -f(\mathbf{Z}_{\mathrm{r}}, \mathbf{Z}_{\mathrm{c}}) + \rho_n \, p(\mathbf{Z}_{\mathrm{r}}, \mathbf{Z}_{\mathrm{c}}), \label{op2} \\
\text{s.t.} \quad & \mathrm{KLD}_{\mathrm{r},t} \geq \underline{A}_t, \quad \forall t, \\
& \mathrm{KLD}_{\mathrm{c},k} \geq \underline{B}_k, \quad \forall k, \\
& \|\mathbf{Z}_{\mathrm{c}}\|_F^2 + \sum_{t=1}^{T} \frac{1}{L} \|\mathbf{Z}_{\mathrm{r},t}\|_F^2 \leq P_\mathrm{T}, \label{eq:auxiliary_const3}\vspace{-0.05 in}
\end{align}
where $p(\mathbf{Z}_{\mathrm{r}}, \mathbf{Z}_{\mathrm{c}})$ is a penalty term that ensures constraints are satisfied. Since the subproblem \eqref{op2} may not have a closed-form solution, it is solved approximately using projected gradient descent. The auxiliary variables are iteratively updated as: $\mathbf{Z}_{\mathrm{r}}^{(n+1)} = \mathbf{Z}_{\mathrm{r}}^{(n)} - \alpha_n \left( \nabla_{\mathbf{Z}_{\mathrm{r}}} f - \rho_n\, \nabla_{\mathbf{Z}_{\mathrm{r}}} p \right)$, and
$\mathbf{Z}_{\mathrm{c}}^{(n+1)} = \mathbf{Z}_{\mathrm{c}}^{(n)} - \alpha_n \left( \nabla_{\mathbf{Z}_{\mathrm{c}}} f - \rho_n\, \nabla_{\mathbf{Z}_{\mathrm{c}}} p \right)$, where $\rho_n$ is the penalty parameter at iteration $n$, and $\alpha_n$ is the step size determined by a backtracking line search. The gradients of $f$ with respect to $\mathbf{Z}_{\mathrm{r}}$ and $\mathbf{Z}_{\mathrm{c}}$ are computed as,
\begin{align}
\vspace{-0.05 in}
\nabla_{\mathbf{Z}_{\mathrm{r}}} f &= \frac{1}{T} \sum_{t=1}^{T} \nabla_{\mathbf{Z}_{\mathrm{r}}} \mathrm{KLD}_{\mathrm{r},t} - \rho (\mathbf{Z}_{\mathrm{r}} - \mathbf{V}_{\mathrm{r}}), \\
\nabla_{\mathbf{Z}_{\mathrm{c}}} f &= \frac{1}{K} \sum_{k=1}^{K} \nabla_{\mathbf{Z}_{\mathrm{c}}} \mathrm{KLD}_{\mathrm{c},k} - \rho (\mathbf{Z}_{\mathrm{c}} - \mathbf{V}_{\mathrm{c}}).\vspace{-0.05 in}
\end{align}
The closed-form expressions for $\nabla_{\mathbf{Z}_{\mathrm{r}}} \mathrm{KLD}_{\mathrm{r},t}$ and $\nabla_{\mathbf{Z}_{\mathrm{c}}} \mathrm{KLD}_{\mathrm{c},k}$ are given by,
\begin{align}\vspace{-0.05 in}
\nabla_{\mathbf{Z}_{\mathrm{r}}} \mathrm{KLD}_{\mathrm{r},t} &= 2\mathbf{Z}_{\mathrm{r},t}\left(\mathbf{R}_{t}^{-1} - \mathbf{R}_{t}^{-1} \mathbf{R}_{2,t} \mathbf{R}_{t}^{-1}\right), \label{eq:radar_gradient} \\
\nabla_{\mathbf{Z}_{\mathrm{c}}} \mathrm{KLD}_{\mathrm{c},k} &= \frac{2\lambda d_{\mathrm{c},k}^{-\zeta}}{M(M-1)\ln 2}\Bigg[\frac{\mathbf{h}_{k}\mathbf{h}_{k}^T \mathbf{Z}_{\mathrm{c},k}}{\sigma_\eta^2 + \sum_{j\neq k} |\mathbf{h}_{k}^T \mathbf{Z}_{\mathrm{c},j}|^2} \nonumber \\
&\quad - \sum_{i\neq k} \frac{\mathbf{h}_{k}\mathbf{h}_{k}^T \mathbf{Z}_{\mathrm{c},i}}{\left(\sigma_\eta^2 + \sum_{j\neq k} |\mathbf{h}_{k}^T \mathbf{Z}_{\mathrm{c},j}|^2\right)^2}\Bigg]. \label{eq:comm_gradient}\vspace{-0.05 in}
\end{align}
After each gradient-based update, the auxiliary variables are projected onto the feasible set defined by the power constraint as, $\mathbf{Z}_{\mathrm{r}}^{(n+1)} = \mathcal{P}\left( \mathbf{Z}_{\mathrm{r}}^{(n+1)} \right)$, and $\mathbf{Z}_{\mathrm{c}}^{(n+1)} = \mathcal{P}\left( \mathbf{Z}_{\mathrm{c}}^{(n+1)} \right)$, where $\mathcal{P}(\cdot)$ denotes the projection operator ensuring that \eqref{eq:auxiliary_const3} is satisfied. This iterative procedure continues until convergence criteria, such as primal and dual residual thresholds, are met.
\begin{figure}[t]
\vspace{-0.3 in}
\begin{algorithm}[H]
\caption{ADMM for Unified ISAC Waveform optimisation}
\label{alg_admm}
\begin{algorithmic}[1]
\REQUIRE Initial variables $\mathbf{W}_{\mathrm{r}}^{(0)}$, $\mathbf{W}_{\mathrm{c}}^{(0)}$, $\mathbf{Z}_{\mathrm{r}}^{(0)}$, $\mathbf{Z}_{\mathrm{c}}^{(0)}$, $\mathbf{U}_{\mathrm{r}}^{(0)}$, $\mathbf{U}_{\mathrm{c}}^{(0)}$, parameters $\rho$, $\alpha_0$, $\rho_0$, tolerance $\varepsilon$, max iterations $\mathrm{max\_iter}$
\ENSURE Optimised waveforms $\mathbf{W}_{\mathrm{r}}^*$, $\mathbf{W}_{\mathrm{c}}^*$
\STATE Initialise $n = 0$
\WHILE{$n < \mathrm{max\_iter}$}
\STATE Update $\mathbf{W}_{\mathrm{r}}^{(n+1)}$ using \eqref{up1}
\STATE Update $\mathbf{W}_{\mathrm{c}}^{(n+1)}$ using \eqref{up2}
\STATE Solve optimisation sub-problem \eqref{op2} to update $\mathbf{Z}_{\mathrm{r}}^{(n+1)}$, $\mathbf{Z}_{\mathrm{c}}^{(n+1)}$ using the projected gradient descent
\STATE Project $\mathbf{Z}_{\mathrm{r}}^{(n+1)}$, $\mathbf{Z}_{\mathrm{c}}^{(n+1)}$ onto feasible set
\STATE Update $\mathbf{U}_{\mathrm{r}}^{(n+1)}$ using \eqref{up4}
\STATE Update $\mathbf{U}_{\mathrm{c}}^{(n+1)}$ using \eqref{up5}
\IF{$\left\| \mathbf{W}_{\mathrm{r}}^{(n+1)} - \mathbf{Z}_{\mathrm{r}}^{(n+1)} \right\|_F < \varepsilon$ \textbf{and} $\left\| \mathbf{W}_{\mathrm{c}}^{(n+1)} - \mathbf{Z}_{\mathrm{c}}^{(n+1)} \right\|_F < \varepsilon$}
\STATE \textbf{break}
\ENDIF
\STATE $n = n + 1$
\ENDWHILE
\RETURN $\mathbf{W}_{\mathrm{r}}^* = \mathbf{W}_{\mathrm{r}}^{(n)}$, $\mathbf{W}_{\mathrm{c}}^* = \mathbf{W}_{\mathrm{c}}^{(n)}$
\end{algorithmic}
\end{algorithm}
\vspace{-0.3in}
\end{figure}
\vspace{-0.05in}
\section{Complexity Analysis \label{Sec-ca}}
\subsubsection{Projected Gradient Method with Penalty}
The projected gradient method with adaptive penalties for radar waveform optimisation, defined in Algorithm \ref{alg1}, has an estimated per-iteration complexity as \cite{armijo1966,bertsekas1999nonlinear},
\vspace{-0.05 in}
\begin{equation}
\mathcal{O}(I \times (E + n^3 + (k+1) \times E + m \times n^2)),\vspace{-0.05 in}
\end{equation}
where $I$ is the number of iterations, $E$ denotes the cost of evaluating the objective function and its gradients, $n$ represents the number of optimisation variables in the tensor $\dot{\mathbf{W}}_\mathrm{r}$, $m$ is the number of constraints, and $k$ is the number of backtracking line search iterations per step. The dominant $n^3$ term arises from matrix operations in the projection step.

\subsubsection{Gradient-assisted IPM} 
For communication waveform optimisation, we utilise a gradient-assisted Interior Point Method as outlined in Algorithm~\ref{alg2}. Its per-iteration complexity is \cite{nocedal2006interior},
\vspace{-0.05 in}
\begin{equation}
\mathcal{O}(I \times (E + n^3 + m \times n^2)),\vspace{-0.05 in}
\end{equation}
where $n$ denotes the number of optimisation variables in the matrix $\mathbf{W}_\mathrm{c}$. Again, while the dominant $n^3$ term reflects matrix operations.
\subsubsection{ADMM} 
The ADMM algorithm for unified optimisation defined in Algorithm \ref{alg_admm}, has per-iteration complexity as \cite{8186925},
\vspace{-0.05 in}
\begin{equation}
\mathcal{O}(I \times (E + n^3 + J \times E + m \times n^2)),\vspace{-0.05 in}
\end{equation}
where $n$ represents the combined number of optimisation variables, and $J$ is the number of inner projected gradient iterations per ADMM step. The dominant $n^3$ term arises from matrix operations in the projection steps.
\vspace{-0.12 in}
\section{Numerical Results\label{Sec-results}}

In this section, we present simulation results for our conventional benchmarks and the three previously introduced optimisation techniques. The total transmit power is fixed at \(P_\mathrm{T}=1\), and QPSK modulation is used throughout this section. The base station has \(N=20\) antennas, with \(L=100\) snapshots, a maximum of \(T=3\) potential radar targets and \(K=3\) UEs are deployed. A pathloss exponent of \(\zeta=3\) is considered to model large-scale fading. The distances from the BS to the UEs are $d_{\mathrm{c},k}=\{150, 210, 100\} \forall k\leq K$ meters and to the targets are $d_{\mathrm{r},t}=\{100, 115, 95\}\forall t \leq T$ meters. The channel variance is fixed at \(\sigma_h^2=1\), the probability of false alarm $P_{FA}$ is set to $10^{-4}$, $\overline{C}_{k}$ is set to $10^-2$ which is approximately equivalent to the $\mathrm{KLD_c}$ of $10$ bits, $N_o$ is the noise power spectral density, and $P_\mathrm{R}$ is the total received power. 
\vspace{-0.1 in}
\subsection{Conventional benchmark}

The radar subsystem employs a CIC matrix design, with \(\mathbf{R}_{w}= \mathbf{I}_{N}\), while the communication subsystem utilises the previously introduced ZF.
\begin{figure}[!ht]
\centering
\vspace{-0.07in}
\includegraphics[width=3.3in]{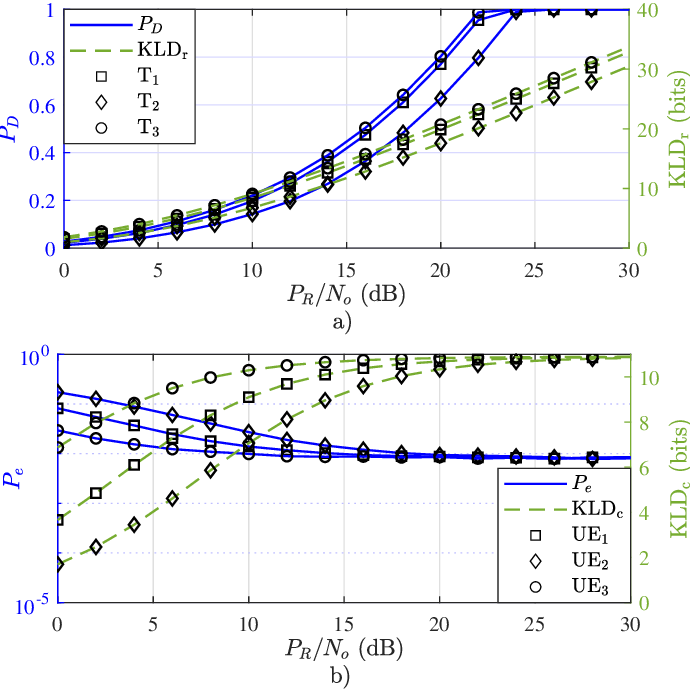} 
\vspace{-0.1in}
\caption{ISAC performance at \(P_\mathrm{r}=0.5\). a) Radar subsystem CIC design performance: $P_D$ and $\mathrm{KLD}_\mathrm{r}$ vs $P_\mathrm{R}/N_o$. b) Communication subsystem ZF precoding performance: BER and \(\mathrm{KLD}_\mathrm{c}\) vs \(P_\mathrm{R}/{N}_o\).}
\label{fig:1}
\vspace{-0.16 in}
\end{figure}

Fig.\ref{fig:1} presents the ISAC system performance, analysing both its radar and communication subsystems at $P_\mathrm{c}=0.5$. The analysis encompasses the radar subsystem's target detection capabilities and the communication subsystem's data transmission quality. Fig.\ref{fig:1}.a presents the performance of the radar subsystem, illustrated in terms of the $\mathrm{KLD_r}$ and the probability of detection $P_D$ against $P_\mathrm{R}/N_o$ for three potential targets $\mathrm{T}_1$, $\mathrm{T}_2$, and $\mathrm{T}_3$. The curves for all three targets exhibit a consistent upward trend with increasing $P_\mathrm{R}/N_o$, indicating that higher $P_\mathrm{R}/N_o$ levels enhance the radar subsystem’s capability to distinguish and detect targets. It can be observed from the figure that at $P_\mathrm{R}/N_o=10$~dB, the $\mathrm{KLD_\mathrm{r}}$ values for $\mathrm{T}_1$, $\mathrm{T}_2$, and $\mathrm{T}_3$ are approximately $\{8.43,\, 6.77,\, 9.1\}$ bits, respectively, while the corresponding detection probabilities $P_D$ reach $\{0.197,\, 0.143,\, 0.221\}$. By the time $P_\mathrm{R}/N_o$ increases to $20$~dB, the $\mathrm{KLD_\mathrm{r}}$ improves to around $\{19.87,\, 17.59,\, 20.71\}$ bits, and $P_D$ increases to $\{0.771,\, 0.627,\, 0.803\}$. Throughout these $P_\mathrm{R}/N_o$ values, $\mathrm{T}_3$ remains the most detectable target, consistently demonstrating the highest $\mathrm{KLD_\mathrm{r}}$ and $P_D$, followed by $\mathrm{T}_1$ and then $\mathrm{T}_2$. The consistent rise in both $\mathrm{KLD_\mathrm{r}}$ and $P_D$ with increasing $P_\mathrm{R}/N_o$ demonstrates the strong correlation between the two measures, confirming the validity of utilising $\mathrm{KLD_\mathrm{r}}$ as a reliable measure of radar performance, as well as a reliable design tool. Higher $\mathrm{KLD_\mathrm{r}}$ values are directly translated to improved detection probabilities, reinforcing that $\mathrm{KLD_\mathrm{r}}$ effectively captures the system's ability to distinguish targets. Fig.\ref{fig:1}.b shows the performance of the communication subsystem under the ZF precoding scheme, where the $\mathrm{KLD_\mathrm{c}}$ and the BER are plotted versus $P_\mathrm{R}/N_o$ for three single-antenna UEs. It can be seen that, as $P_\mathrm{R}/N_o$ increases, the KLD for all UEs rises, reflecting enhanced signal distinguishability. Correspondingly, the BER decreases, indicating better communication quality. Around $10$~dB, however, both an upper bound in KLD and an error floor in BER become evident, highlighting the persistent radar interference as the dominant performance limiting factor once IUI has been largely cancelled by ZF at $P_\mathrm{c}=0.5$. Thereby, this validates the use of KLD as an effective tool for characterising and designing communication systems since higher KLD values systematically align with lower BER.

\begin{figure}[!ht]
\centering
\vspace{-0.07in}
\includegraphics[width=3.3in]{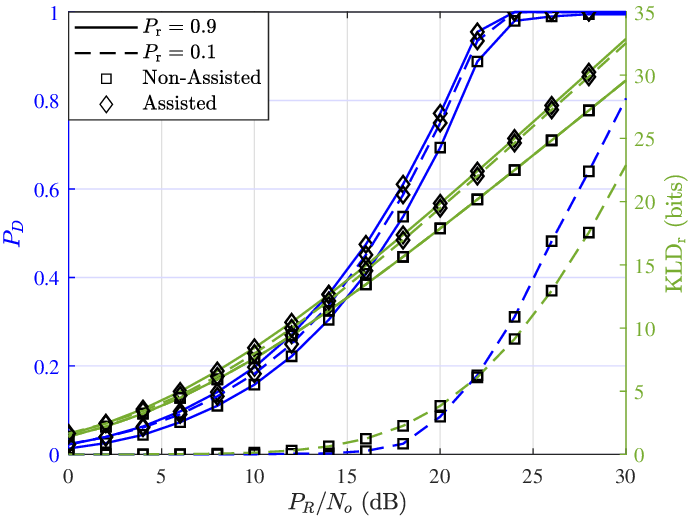} 
\vspace{-0.1in}
\caption{Radar subsystem CIC design performance comparison between communication-assisted and non-assisted configurations at $P_\mathrm{r}=\{0.1, 0.9\}$: $P_D$ and $\mathrm{KLD}_\mathrm{r}$ vs $P_\mathrm{R}/N_o$.}
\label{fig:comm_assist}
\vspace{-0.12 in}
\end{figure}

Fig.~\ref{fig:comm_assist} compares communication-assisted and non-assisted radar configurations for $P_\mathrm{r}=0.9$ and $P_\mathrm{r}=0.1$ for target $\mathrm{T}_1$. The assisted configuration exploits communication signals as an additional information source for radar detection, while the non-assisted baseline considers only dedicated radar signals. Importantly, the communication subsystem performance remains unaffected by the choice of assisted or non-assisted radar operation, as the exploitation occurs solely at the radar receiver through processing the reflected signals. For $P_\mathrm{r}=0.1$, the communication-assisted configuration demonstrates substantial performance gains. At $P_\mathrm{R}/N_o=20$~dB, the assisted system achieves $\mathrm{KLD}_\mathrm{r} = 19.489$ bits compared to $3.842$ bits for the non-assisted case, representing a remarkable improvement. The corresponding detection probabilities reach $P_D = 0.749$ and $0.086$, respectively. This significant enhancement stems from the high communication power $P_\mathrm{c}=0.9$ providing substantial additional signal energy that, when reflected from target $\mathrm{T}_1$, contributes to the detection process. At $P_\mathrm{r}=0.9$, both configurations show improved performance, with the assisted system maintaining a consistent advantage. At $P_\mathrm{R}/N_o=20$~dB, the assisted configuration achieves $\mathrm{KLD}_\mathrm{r} = 19.873$ bits versus $17.795$ bits for the non-assisted scenario, yielding a modest but good improvement. The detection probabilities are $P_D = 0.771$ and $0.694$, respectively. The reduced relative gain is expected as the dedicated radar signal already provides substantial detection capability, diminishing the proportional contribution from communication signals. The consistent performance enhancement across all SNR levels confirms that our communication-assisted approach effectively exploits the inherent synergy between communication and radar signals.

\vspace{-0.09 in}
\subsection{Radar Waveform KLD-based Optimisation}\label{sec-b}
The radar waveform KLD-based optimisation algorithm is set with a maximum number of iterations of $\mathrm{max\_iter} = 1000$, convergence tolerance $\epsilon = 10^{-6}$, initial penalty $\rho_0 = 1$, penalty increase $\gamma = 1.5$, and initial step size $\alpha_0 = 0.1$. The KLD lower bounds $\underline{A}_t$ and $\underline{B}_k$ for all targets and UEs are 10 bits.
\begin{figure}[!ht]
\centering
\vspace{-0.05in}
\includegraphics[width=3.3in]{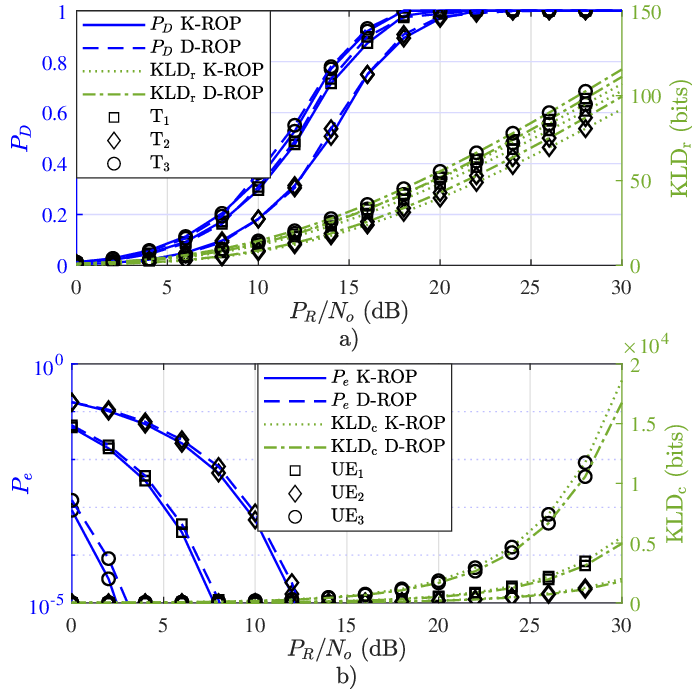} 
\vspace{-0.18in}
\caption{ ISAC performance at \(P_\mathrm{r}=0.5\) using two radar waveform optimisations: K-ROP and D-ROP. a) Radar subsystem: $P_D$ and $\mathrm{KLD}_\mathrm{r}$ vs $P_\mathrm{R}/N_o$. b) Communication subsystem: BER and \(\mathrm{KLD}_\mathrm{c}\) vs \(P_\mathrm{R}/{N}_o\).}
\label{fig:2}
\vspace{-0.15 in}
\end{figure}

Fig. \ref{fig:2} illustrates the performance of both subsystems using both KLD-based radar waveform optimisation (K-ROP) and detection probability radar waveform optimisation (D-ROP) techniques. In Fig. \ref{fig:2}.a, the achievable $\mathrm{KLD_\mathrm{r}}$ and detection probability $P_D$ versus $P_\mathrm{R}/N_o$ are shown for each target. As can be depicted from the figure, as $P_\mathrm{R}/N_o$ increases, both measures improve for all targets, indicating enhanced radar performance. The results in this figure also show that a slight improvement can be achieved using D-ROP compared to K-ROP, at the expense of slightly worse communication performance due to increased interference. Moreover, by comparing the results here with the ZF-based design in Fig. \ref{fig:1}, it can be observed that a significant gain can be achieved using K-ROP and D-ROP. For example, at $P_\mathrm{R}/{N}_o= 18$ dB, the achievable $P_{D}$ is $\{98, 89.2, 99.9\}\%$ for $\mathrm{T}_1$, $\mathrm{T}_2$, and $\mathrm{T}_3$ using K-ROP, compared to $\{61, 48.3, 64.2\}\%$ in the non-optimised ZF in Fig. \ref{fig:1}, while D-ROP achieves $\{97.4, 90.7, 99.99\}\%$, demonstrating improved detection capabilities by both optimisation techniques with D-ROP showing a slight improvement, at the cost of higher complexity. For K-ROP, $\mathrm{T}_3$ consistently achieves the highest $\mathrm{KLD_\mathrm{r}}$, followed by $\mathrm{T}_1$ and $\mathrm{T}_2$, reflecting their distances from the BS. At $P_\mathrm{R}/N_o = 10$ dB, the $\{\mathrm{KLD}_{\mathrm{r},1},\mathrm{KLD}_{\mathrm{r},2},\mathrm{KLD}_{\mathrm{r},3}\}$ values are approximately $\{11.78, 8.06, 13.38\}$ bits, increasing to $\{48.06, 39.68, 51.31\}$ bits at $P_\mathrm{R}/N_o = 20$ dB. While for D-ROP, $\{\mathrm{KLD}_{\mathrm{r},1},\mathrm{KLD}_{\mathrm{r},2},\mathrm{KLD}_{\mathrm{r},3}\}$ approximately $\{12.73, 8.7, 14.4\}$ bits at $P_\mathrm{R}/N_o = 10$ dB, increasing to $\{51.2, 42.9, 55.41\}$ bits at $P_\mathrm{R}/N_o = 20$ dB. The rate of $\mathrm{KLD_\mathrm{r}}$ improvement after 15 dB suggests a near-linear relationship with $P_\mathrm{R}/N_o$ for both techniques, indicating consistent performance enhancement in the radar system. The continued increase in $\mathrm{KLD}_r$ even as $P_D$ approaches unity demonstrates a key advantage of the KLD. While probability metrics saturate at their bounds, making it difficult to distinguish between detection probabilities closer to $1$, the unbounded KLD clearly differentiates these performance levels. This sensitivity is crucial for high-reliability applications and provides better numerical properties for optimisation algorithms.
 
Fig. \ref{fig:2}.b illustrates the communication subsystem performance, showing $\mathrm{KLD_\mathrm{c}}$ and BER versus $P_\mathrm{R}/N_o$. For K-ROP, $\mathrm{UE}_3$ consistently achieves the highest $\mathrm{KLD_\mathrm{c}}$, followed by $\mathrm{UE}_1$ and $\mathrm{UE}_2$. At $P_\mathrm{R}/N_o = 10$ dB, the values $\{\mathrm{KLD}_{\mathrm{c},1},\mathrm{KLD}_{\mathrm{c},2},\mathrm{KLD}_{\mathrm{c},3}\}$ are approximately $\{55.14, 20.09, 186.09\}$ for $\mathrm{UE}_1$, $\mathrm{UE}_2$, and $\mathrm{UE}_3$, respectively, increasing to $\{551.38, 200.94, 1860.91\}$ at $P_\mathrm{R}/N_o = 20$ dB. D-ROP demonstrates similar trends with slightly lower performance, achieving $\{\mathrm{KLD}_{\mathrm{c},1},\mathrm{KLD}_{\mathrm{c},2},\mathrm{KLD}_{\mathrm{c},3}\}$ values of $\{49.62, 18.07, 167.48\}$ bits at $P_\mathrm{R}/N_o = 10$ dB and $\{496.2, 180.84, 1674.79\}$ bits at $P_\mathrm{R}/N_o = 20$ dB. The sharp increase in $\mathrm{KLD_\mathrm{c}}$ beyond $8$ dB for both techniques suggests they have reached their optimisation feasibility points, significantly enhancing the communication performance as the radar interference is effectively reduced. The BER performance shows that K-ROP achieves $P_e$ below $10^{-5}$ by $13$ dB, while D-ROP reaches this threshold at $14$ dB, both significantly outperforming the non-optimised ZF case where $P_e$ saturates at $8 \times 10^{-3}$ after 15 dB as depicted in Fig. \ref{fig:1}. These results highlight the intricate relationship between radar detection capabilities and communication performance, with K-ROP demonstrating slightly superior performance over D-ROP.

\vspace{-0.13 in}
\subsection{Communication Waveform KLD-based Optimisation \label{sec-c}}

The communication waveform KLD-based optimisation is configured similarly to the radar algorithm, with a maximum iteration limit of $\mathrm{max\_iter} = 1000$ and a convergence tolerance of $\epsilon = 10^{-6}$. The barrier parameter is initialised at $\tilde{\mu_0} = 1$ and the barrier reduction factor at $\tilde{\gamma} = 0.1$. The KLD lower bounds for all targets and UEs, $\underline{A}_t$ and $\underline{B}_k$, are set to 10 bits.
\begin{figure}[!ht]
\centering
\vspace{-0.05in}
\includegraphics[width=3.3in]{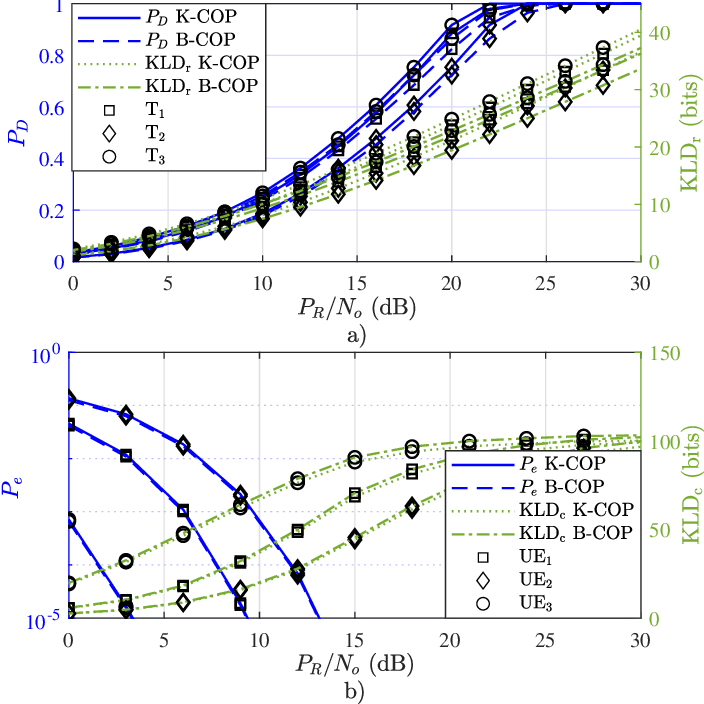} 
\vspace{-0.12in}
\caption{ISAC performance at \(P_\mathrm{r}=0.5\) using two communication waveform optimisations: K-COP and B-COP. a) Radar subsystem: $P_D$ and $\mathrm{KLD}_\mathrm{r}$ vs $P_\mathrm{R}/N_o$. b) Communication subsystem: BER and \(\mathrm{KLD}_\mathrm{c}\) vs \(P_\mathrm{R}/{N}_o\).}
\label{fig:3}
\vspace{-0.10 in}
\end{figure}

Fig. \ref{fig:3} illustrates the performance of both subsystems using both KLD-based communication waveform optimisation (K-COP) and BER-based communication waveform optimisation (B-COP) techniques. In Fig. \ref{fig:3}.a, the achievable $\mathrm{KLD_\mathrm{r}}$ and detection probability $P_D$ versus $P_\mathrm{R}/N_o$ are shown for each target. For K-COP, as $P_\mathrm{R}/N_o$ increases, $\mathrm{KLD_\mathrm{r}}$ improves across all targets, with target $\mathrm{T}_3$ consistently achieving the highest $\mathrm{KLD_\mathrm{r}}$, followed by $\mathrm{T}_2$ and $\mathrm{T}_1$. At $P_\mathrm{R}/N_o = 10$ dB, the $\{\mathrm{KLD}_{\mathrm{r},1},\mathrm{KLD}_{\mathrm{r},2},\mathrm{KLD}_{\mathrm{r},3}\}$ values are approximately $\{10.12, 8.13, 10.91\}$ bits, increasing to $\{23.84, 21.1, 24.85\}$ bits at $P_\mathrm{R}/N_o = 20$ dB. B-COP shows slightly lower $\mathrm{KLD_\mathrm{r}}$ performance, with values $\{9.31, 7.48, 10.04\}$ bits at $P_\mathrm{R}/N_o = 10$ dB, rising to $\{21.93, 19.42, 22.86\}$ bits at $P_\mathrm{R}/N_o = 20$ dB. The linear improvement in $\mathrm{KLD_\mathrm{r}}$ beyond $15$ dB for both techniques indicates stable radar performance despite the suboptimal waveform. At $P_\mathrm{R}/N_o= 18$ dB, the $P_{D}$ values for K-COP are $\{72.2, 60.7, 75.4\}\%$ for $\mathrm{T}_1$, $\mathrm{T}_2$, and $\mathrm{T}_3$, respectively, while B-COP achieves $\{68.5, 58.2, 72.3\}\%$, compared to $\{61, 48.3, 64.2\}\%$ for the non-optimised ZF shown in Fig. \ref{fig:1}. Both techniques demonstrate improved detection capabilities over the non-optimised scenario but remain below the radar waveform optimisation performance achieved using K-ROP in Fig. \ref{fig:2}.

Fig. \ref{fig:3}.b illustrates the communication subsystem performance, showing $\mathrm{KLD_\mathrm{c}}$ and BER versus $P_\mathrm{R}/N_o$. For K-COP, $\mathrm{UE}_3$ achieves the highest $\mathrm{KLD_\mathrm{c}}$, followed by $\mathrm{UE}_2$ and $\mathrm{UE}_1$. For instance, at $P_\mathrm{R}/N_o = 10$ dB, the $\{\mathrm{KLD}_{\mathrm{c},1},\mathrm{KLD}_{\mathrm{c},2},\mathrm{KLD}_{\mathrm{c},3}\}$ values are approximately $\{36.94, 19.55, 66.52\}$ bits, rising to $\{87.06, 71.15, 95.94\}$ bits at $P_\mathrm{R}/N_o = 20$ dB. Interestingly, B-COP demonstrates marginally better communication performance. The BER performance reflects this trend, with B-COP achieving slightly better performance than K-COP. In the non-optimised ZF scenario, $P_e$ saturates at $8 \times 10^{-3}$ after $15$ dB (Fig. \ref{fig:1}), while with K-COP optimisation, $P_e$ drops below $10^{-5}$ by $14$ dB across all UEs. B-COP achieves this threshold marginally earlier at $13.5$ dB. Both optimisation techniques provide significant performance gains over the non-optimised ZF, though these gains are less pronounced than those achieved through radar waveform optimisation in Sec. \ref{sec-b} due to radar interference, where K-COP's fixed omnidirectional radar CIC matrix design creates an interference ceiling that limits achievable communication performance regardless of precoding sophistication. The communication performance shows less improvement at higher $P_\mathrm{R}/N_o$ due to the need to balance radar subsystem performance with communication enhancements.

\vspace{-0.12in}
\subsection{Integrated Waveform KLD-based Optimisation for ISAC \label{sec-d} \label{sec-IOP}}
The integrated waveform optimisation algorithm is set with a maximum of $\mathrm{max\_iter} = 1000$, convergence tolerance $\epsilon = 10^{-6}$, initial penalty $\rho_0 = 1$, $\rho=1$, penalty increase $\gamma = 1.5$, and initial step size $\alpha_0 = 0.1$. The KLD lower bounds $\underline{A}_t$ and $\underline{B}_k$ for all targets and UEs are 10 bits.

\begin{figure}[!ht]
\centering
\vspace{-0.05in}
\includegraphics[width=3.3in]{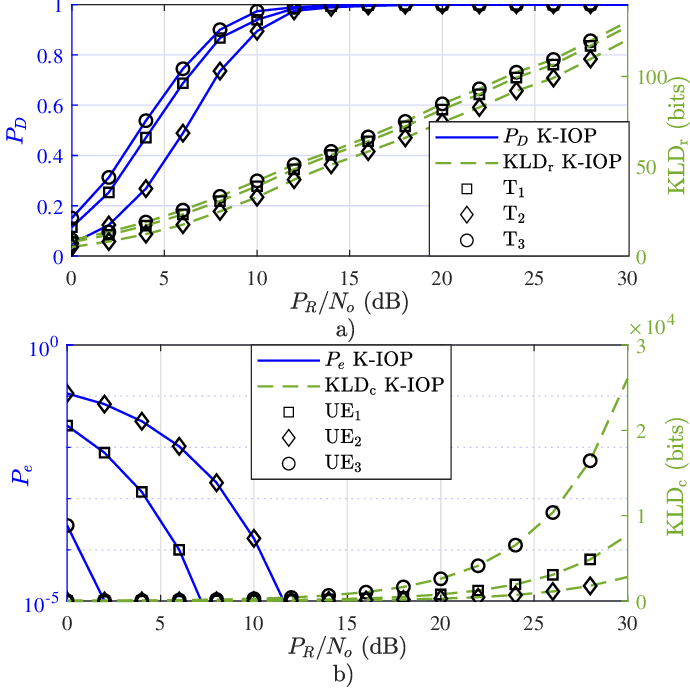} 
\vspace{-0.12in}
\caption{ISAC performance using the integrated waveform optimisation. a) Radar subsystem performance: $P_D$ and $\mathrm{KLD}_\mathrm{r}$ vs $P_\mathrm{R}/N_o$. b) Communication subsystem performance: BER and \(\mathrm{KLD}_\mathrm{c}\) vs \(P_\mathrm{R}/{N}_o\).}
\label{fig:comb1}
\vspace{-0.10 in}
\end{figure}

Fig. \ref{fig:comb1} illustrates the ISAC system performance under the KLD-based integrated waveform optimisation (K-IOP). 
In Fig. \ref{fig:comb1}.a, the radar subsystem performance is shown in terms of detection probability \(P_D\) and \(\mathrm{KLD}_\mathrm{r}\) for each target \(\mathrm{T}_1\), \(\mathrm{T}_2\), and \(\mathrm{T}_3\). As \(\mathit{P}_\mathrm{R}/N_o\) increases from 0 to 30\,dB, both \(P_D\) and \(\mathrm{KLD}_\mathrm{r}\) improve for all targets, confirming the effectiveness of K-IOP. For instance, at \(\mathit{P}_\mathrm{R}/N_o = 10\) dB, the achievable \(\mathrm{KLD}_{\mathrm{r}}\) values for \(\{\mathrm{T}_1,\mathrm{T}_2,\mathrm{T}_3\}\) are approximately \(\{38.87, 32.77, 41.85\}\) bits, with corresponding detection probabilities \(\{0.94, 0.89, 0.97\}\). Performance continues to rise at higher \(\mathit{P}_\mathrm{R}/N_o\), achieving near-unity detection across all targets once \(\mathit{P}_\mathrm{R}/N_o\) exceeds 22\,dB.

In Fig. \ref{fig:comb1}.b, the communication subsystem performance is depicted via the BER and \(\mathrm{KLD}_\mathrm{c}\) for each UE \(\mathrm{UE}_1\), \(\mathrm{UE}_2\), and \(\mathrm{UE}_3\). 
Similar to the radar trends, higher \(\mathit{P}_\mathrm{R}/N_o\) values yield a marked improvement in both \(\mathrm{KLD}_\mathrm{c}\) and BER. 
For example, at \(\mathit{P}_\mathrm{R}/N_o = 10\) dB, \(\mathrm{KLD}_\mathrm{c}\) values of \(\{77.19, 28.13, 260.53\}\) bits are observed for \(\{\mathrm{UE}_1,\mathrm{UE}_2,\mathrm{UE}_3\}\), with corresponding BERs of approximately \(\{5.1\times 10^{-9}, 1.7\times 10^{-4},  \lesssim 10^{-10}\}\). Beyond 14-16\,dB, the BER for all UEs drops rapidly-- often reaching effectively error-free communication-- and \(\mathrm{KLD}_\mathrm{c}\) grows significantly as the interference from the radar subsystem is jointly minimised. Comparing these K-IOP results with the non-optimised ZF baseline (cf. Fig.\ref{fig:1}) or the separate (radar-only or communication-only) waveform approaches in Sec.~\ref{sec-b}--\ref{sec-c}, it is evident that the integrated design offers substantially improved performance simultaneously for radar detection and communication reliability. This highlights the advantages of a truly integrated ISAC paradigm, where waveforms are optimally tailored to balance both subsystems' requirements.

\begin{figure}[!ht]
\centering
\vspace{-0.05in}
\includegraphics[width=3.3in]{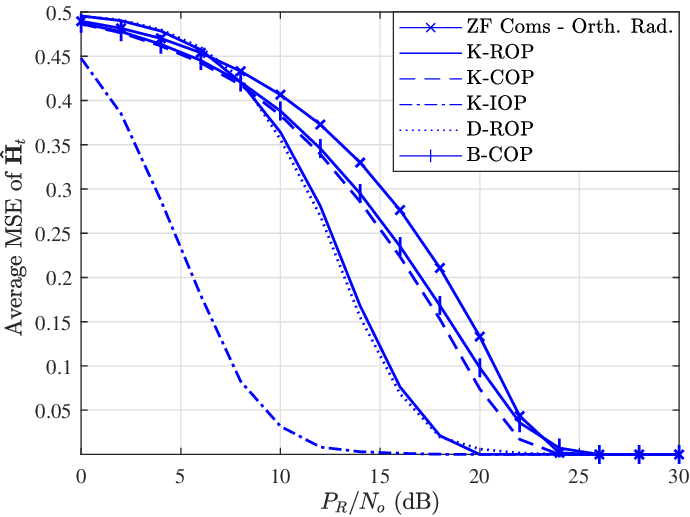} 
\vspace{-0.1in}
    \caption{Average MSE of the target response matrix versus $P_\mathrm{R}/N_o$.}
\label{fig:6}
\vspace{-0.1 in}
\end{figure}
Fig. \ref{fig:6} presents the average MSE of the estimated target response matrix $\mathbf{\hat{H}}_t$ averaged for all targets versus $P_\mathrm{R}/N_o$ for all optimisation techniques. At low $P_\mathrm{R}/N_o$ values ($0$-$10$ dB), K-IOP immediately achieves lower MSE than all of the other techniques. As $P_\mathrm{R}/N_o$ increases beyond $10$ dB, the techniques begin to show more distinct performance. K-ROP and D-ROP demonstrate faster MSE decay, particularly in the $15$-$20$ dB range, while K-COP and B-COP exhibit a steadier improvement but converge slightly slower. The K-IOP continues to outperform all separate waveform approaches at all values of $P_\mathrm{R}/N_o$, reaching near-zero MSE at lower \(P_\mathrm{R}/N_o\) values. Eventually, all techniques attain negligible MSE at sufficiently high $P_\mathrm{R}/N_o$ levels. This MSE performance strongly correlates with the detection capabilities in Figs.~\ref{fig:3}.a, \ref{fig:2}.a, and \ref{fig:comb1}, where more reliable detections yield more accurate target-response estimates under $\mathcal{H}_1$. Conversely, missed detections or false alarms degrade the estimation by introducing noise-based estimates.
\begin{figure}[!ht]
\centering
\vspace{-0.1in}
\includegraphics[width=3.3in]{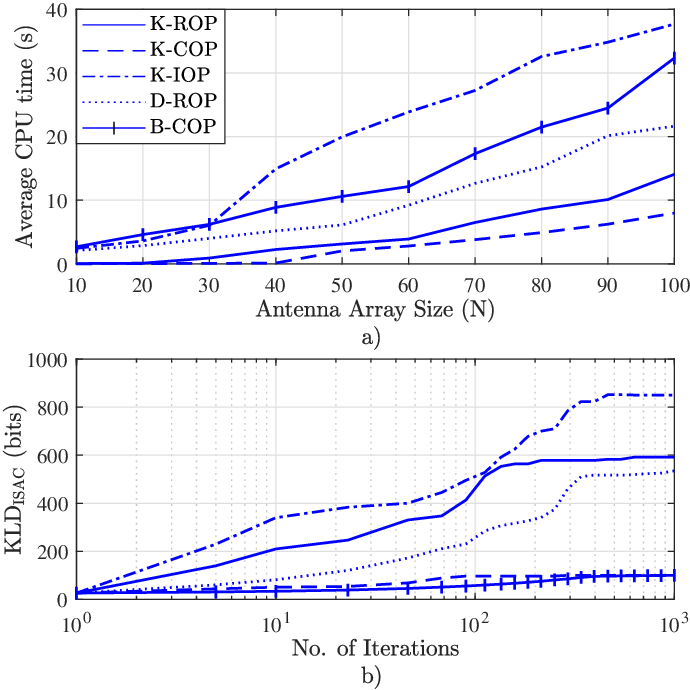} 
\vspace{-0.1in}
    \caption{a) Average CPU time (seconds) vs. antenna array size for both optimisation techniques. b)$\mathrm{KLD_{ISAC}}$ convergence vs. Number of iterations.}
\label{fig:5}
\vspace{-0.1 in}
\end{figure}
Fig. \ref{fig:5}.a presents the average CPU time versus antenna array size $N$ for all optimisation techniques. The K-COP demonstrates the lowest computational cost, increasing from $0.04$s at $N=10$ to $7.98$s at $N=100$. On the other hand, K-ROP exhibits moderate requirements, ranging from $0.001$s to $14.07$s across the same $N$ range, and D-ROP shows higher CPU time growth, reaching $21.63$s at $N=100$. The communication-oriented B-COP requires $32.35$s at $N=100$, while K-IOP incurs the highest computational load at $37.69$s. These results confirm that while the radar-only optimisations maintain relatively low complexity, the joint K-IOP technique and B-COP require additional processing time, with K-IOP achieving superior ISAC performance despite its higher computational cost. Fig. \ref{fig:5}.b illustrates the convergence behaviour of the $\mathrm{KLD_{ISAC}}$ metric versus the iteration count at $P_\mathrm{R}/N_o=18$ dB. As can be observed from the figure, K-IOP demonstrates the most rapid convergence, achieving substantial gains within the first $100$ iterations and reaching $74\%$ of its final value by iteration $158$. K-ROP also shows efficient convergence, stabilising at $95\%$ of its final value by iteration $135$, and D-ROP achieves stabilisation at $95\%$ of its final value by iteration $320$. The communication-focused technique K-COP converges quickly but saturates early, with K-COP reaching $97\%$ of its final value by iteration $90$, while B-COP exhibits a more gradual convergence pattern, requiring approximately $400$ iterations to reach its steady-state value. This analysis reveals that K-IOP efficiently balances rapid convergence with joint radar-communication optimisation, making it particularly suitable for real-time ISAC applications.
\begin{figure}[!ht]
\centering
\vspace{-0.1in}
\includegraphics[width=3.3in]{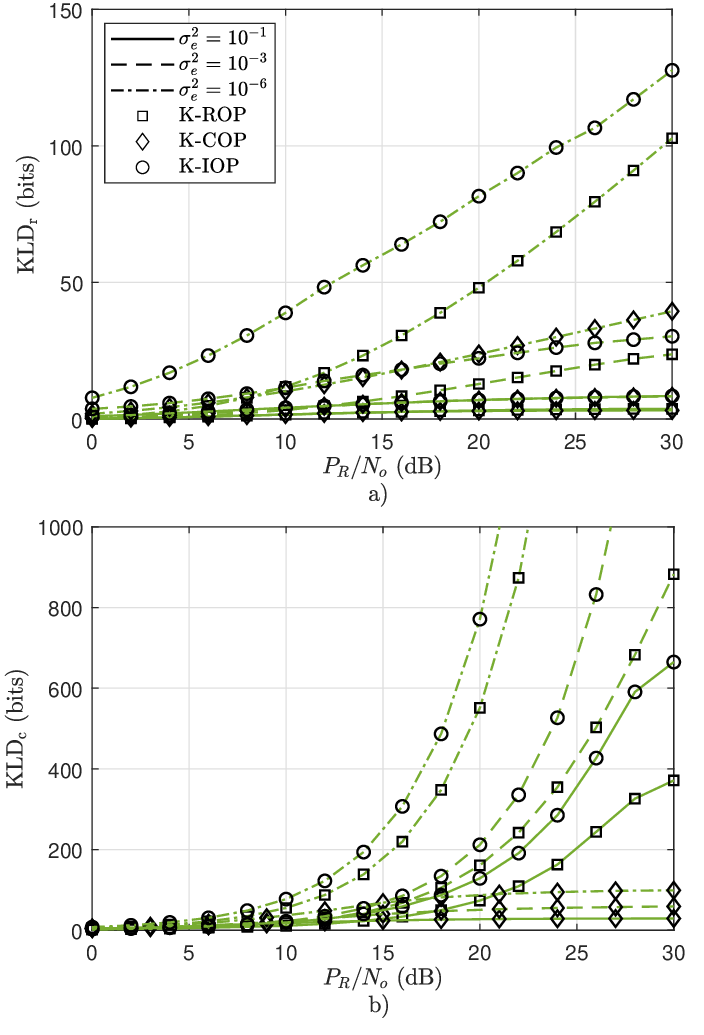} 
\vspace{-0.2in}
    \caption{Achievable KLD under estimation errors for KLD-based optimisations, with $\sigma_e^{2}\!\in\!\{10^{-6},10^{-3},10^{-1}\}$. a) $\mathrm{KLD_r}$ \(P_\mathrm{R}/N_o\). b) $\mathrm{KLD_c}$ vs \(P_\mathrm{R}/N_o\).}
\label{fig:KLD_imperfect_CSI}
\vspace{-0.13 in}
\end{figure}

Fig.~\ref{fig:KLD_imperfect_CSI} illustrates the effect of imperfect channel and radar target-response estimates on the KLD obtained by the proposed optimisation techniques. For clarity, this is shown for $\mathrm{UE}_1$, and $\mathrm{T}_1$. Fig.~\ref{fig:KLD_imperfect_CSI}.a, presents $\mathrm{KLD_r}$ versus \(P_\mathrm{R}/N_o\), and Fig.~\ref{fig:KLD_imperfect_CSI}.b presents $\mathrm{KLD_c}$ versus \(P_\mathrm{R}/N_o\) for three error-variance levels. For each optimisation technique, the quasi-perfect case ($\sigma_e^{2}=10^{-6}$) yields the highest KLD; increasing the error variance to $10^{-3}$ noticeably flattens the curves, and at $10^{-1}$ the gains from all schemes almost disappear. As the error variance rises from $10^{-6}$ to $10^{-1}$, both the radar and communication KLD curves flatten, producing a clear error floor at high $P_\mathrm{R}/N_o$. The observed degradation is therefore a predictable consequence of reduced CSI fidelity, underlining the importance of accurate estimation. 

\vspace{-0.1 in}
\section{Conclusion \label{Sec-con}}
This paper has proposed and validated a novel KLD-based framework for optimising ISAC systems. The three considered optimisation techniques, namely, K-ROP, K-COP, and K-IOP, each demonstrated significant improvements over both the non-optimised scenario and optimisation benchmarks (D-ROP and B-COP). K-ROP and D-ROP yielded substantial enhancements in target detection and communication performance while K-ROP maintained superior computational efficiency compared to D-ROP, validating the analytical tractability advantages of KLD over traditional Q-function-based BER formulations. K-COP showed notable improvements in the communication subsystem with modest radar gains, particularly at lower \(P_\mathrm{R}/N_o\) levels, and demonstrated more efficient optimisation than B-COP due to KLD's tractable gradient expressions. The K-IOP approach effectively unified these advantages, consistently outperforming all separate approaches and benchmarks in detection probability, BER, and estimation accuracy. By jointly designing radar and communication waveforms, K-IOP leverages the synergy between sensing and data transmission to achieve higher overall KLD and more balanced performance. The numerical results revealed that exploiting communication signals within the radar subsystem significantly boosts ISAC capabilities, underscoring the importance of holistic design in shared-antenna deployments. These findings provide valuable insights for next-generation ISAC systems, offering a promising framework that adapts system resources for diverse operating regimes while paving the way for more efficient and versatile wireless communications and sensing technologies. Future work will explore the synergistic interplay between KLD and machine learning in ISAC, extend the framework to emerging architectures including cell-free MIMO and full-duplex systems, and develop practical interference exploitation strategies.

\vspace{-0.09 in}
\appendix
\vspace{-0.03 in}This appendix provides detailed derivations for the gradients used in the optimisation problems.
\vspace{-0.08 in}
\subsection{Gradients for Radar-Waveform Optimisation}
The objective function $f(\mathbf{\dot{W}}_\mathrm{r})$ is given in \eqref{objfunrad}, where the $\mathrm{KLD}_{\mathrm{r},t}$ is defined in \eqref{kldrt}. Therefore, to derive $\nabla f(\mathbf{\dot{W}}_\mathrm{r})$, we first compute $\partial \mathrm{KLD}_{\mathrm{r},t} / \partial \mathbf{W}_{\mathrm{r},t}$. Using the matrix identity $\frac{\partial \ln(\det(\mathbf{X}))}{\partial \mathbf{X}} = (\mathbf{X}^{-1})^T$, and derivative  $\frac{\partial \mathrm{Tr}(\mathbf{X}^{-1}\mathbf{A})}{\partial \mathbf{X}} = -(\mathbf{X}^{-1})^T\mathbf{A}^T(\mathbf{X}^{-1})^T$ we obtain,
\vspace{-0.05 in}
\begin{equation}
\frac{\partial \ln(\det(\mathbf{R}_{2,t}))}{\partial \mathbf{R}_{2,t}}\! =\! \mathbf{R}_{2,t}^{-1}, \quad \frac{\partial \mathrm{Tr}(\mathbf{R}_{2,t}^{-1}\sigma_n^2)}{\partial \mathbf{R}_{2,t}}\!\! =\!\! -\sigma_n^2\mathbf{R}_{2,t}^{-2}.\vspace{-0.05 in}
\label{eq:det_deriv}
\end{equation}
Applying the chain rule to \eqref{kldrt}, and substituting \eqref{eq:det_deriv} we get,
\vspace{-0.03 in}
\begin{equation}
\frac{\partial \mathrm{KLD}_{\mathrm{r},t}}{\partial \mathbf{W}_{\mathrm{r},t}} = \frac{1}{\ln{2}}(\mathbf{R}_{2,t}^{-1} - \sigma_n^2\mathbf{R}_{2,t}^{-2})\cdot\frac{\partial \mathbf{R}_{2,t}}{\partial \mathbf{W}_{\mathrm{r},t}},
\label{eq:kld_deriv}\vspace{-0.05 in}
\end{equation}
where $\frac{\partial \mathbf{R}_{2,t}}{\partial \mathbf{W}_{\mathrm{r},t}} = \frac{d_{\mathrm{r},t}^{\zeta}}{L}\mathbf{H}_t^H\mathbf{H}_t\mathbf{W}_{\mathrm{r},t}$. Substituting this into \eqref{eq:kld_deriv} and simplifying, we obtain,
\vspace{-0.05 in}
\begin{multline}
\!\!\!\!\!\frac{\partial \mathrm{KLD}_{\mathrm{r},t}}{\partial \mathbf{W}_{\!\mathrm{r},t}}\!\! =\! \!\frac{\!d_{\mathrm{r},t}^{\zeta}}{\!L\ln{2}}(\mathbf{H}_t^H\mathbf{R}_{2,t}^{-1}\mathbf{H}_t\!\mathbf{W}_{\!\mathrm{r},t} 
-\sigma_n^2\mathbf{H}_t^H\mathbf{R}_{2,t}^{-2}\mathbf{H}_t\!\mathbf{W}_{\!\mathrm{r},t}\!).\!\!
\label{eq:kld_final_deriv}\vspace{-0.05 in}
\end{multline}
Finally, the gradient of $f$ with respect to $\mathbf{W}_{\mathrm{r},t}$, derived from \eqref{objfunrad} and \eqref{eq:kld_final_deriv}, is,
\vspace{-0.05 in}
\begin{multline}
\!\!\!\!\!\![\nabla \!f(\!\mathbf{\dot{W}}\!_\mathrm{r}\!)]_{i,l,t} \!\!=\!\! \frac{2d_{\mathrm{r},t}^{\zeta}}{TL\!\ln{2}}(\mathbf{H}_t^H\mathbf{R}_{2,t}^{-1}\mathbf{H}_t\!\mathbf{W}_{\!\mathrm{r},t} 
\!- \sigma_n^2\mathbf{R}_{2,t}^{-2}\mathbf{H}_t\!\mathbf{W}_{\!\mathrm{r},t})_i.\!\!\!
\label{eq:final_gradient}\vspace{-0.05 in}
\end{multline}
For $\mathrm{KLD}_{\mathrm{c},k}$ gradient calculations, we consider how changes in $\mathbf{W}_{\mathrm{r}}$ affect $\sigma_{\eta}^2$, where it is defined as follows,
\vspace{-0.06 in}
\begin{equation}
\sigma_{\eta}^2 = d_{\mathrm{c},k}^{\zeta}\:\sigma_h^2\mathrm{Tr}(\frac{1}{L}\sum_{l=1}^{L}{\mathbf{W}_{\mathrm{r},l}\mathbf{W}_{\mathrm{r},l}^H})+\sigma_n^2
\label{eq:sigma_eta}\vspace{-0.05 in}
\end{equation}
The derivative of $\sigma_{\eta}^2$ with respect to $\mathbf{W}_{\mathrm{r},l}$ is,
\vspace{-0.05 in}
\begin{equation}
\frac{\partial \sigma_{\eta}^2}{\partial \mathbf{W}_{\mathrm{r},l}} = \frac{2\:d_{\mathrm{c},k}^{\zeta}\sigma_h^2}{L}\mathbf{W}_{\mathrm{r},l}.
\label{eq:sigma_deriv}\vspace{-0.05 in}
\end{equation}
Computing $\partial \mathrm{KLD}_{\mathrm{c},k} / \partial \sigma_{\eta}^2$,
\vspace{-0.07 in}
\begin{equation}
\frac{\partial \mathrm{KLD}_{\mathrm{c},k}}{\partial \sigma_{\eta}^2} = -\frac{\lambda\:d_{\mathrm{c},k}^{\zeta} p_k (N-K)}{M(M-1) \sigma_{\eta}^2\ln 2}.
\label{eq:kldc_sigma_deriv}\vspace{-0.05 in}
\end{equation}
Utilising the chain rule  and \eqref{eq:sigma_deriv},  \eqref{eq:kldc_sigma_deriv}, we obtain,
\vspace{-0.06 in}
\begin{equation}
\frac{\partial \mathrm{KLD}_{\mathrm{c},k}}{\partial \mathbf{W}_{\mathrm{r},l}} = -\frac{2\lambda\:d_{\mathrm{c},k}^{\zeta} p_k (N-K)\sigma_h^2}{M(M-1) L \sigma_{\eta}^2\ln 2}\mathbf{W}_{\mathrm{r},l}.
\label{eq:kldc_final_deriv}\vspace{-0.05 in}
\end{equation}
Therefore, the gradient of $\mathrm{KLD}_{\mathrm{c},k}$ with respect to $\mathbf{\dot{W}}_\mathrm{r}$ is,
\vspace{-0.05 in}
\begin{equation}
[\nabla \mathrm{KLD}_{\mathrm{c},k}]_{i,l,t} = -\frac{2\lambda\:d_{\mathrm{c},k}^{\zeta} p_k (N-K)\sigma_h^2}{M(M-1) L \sigma_{\eta}^2\ln 2}[\mathbf{\dot{W}}_{\mathrm{r}}]_{i,l,t}.
\label{eq:kldc_gradient}\vspace{-0.06 in}
\end{equation}
Equations \eqref{eq:final_gradient} and \eqref{eq:kldc_gradient} provide the mathematical foundation for the gradients used in our optimisation algorithm.
\vspace{-0.07 in}
\subsection{Gradient of $P_{D,t}$ w.r.t.\ $\mathbf{W}_{\mathrm{r}}$}
\label{app:GradPD}
To enable gradient-based optimisation of $P_{D,t}$, we derive $\nabla_{\mathbf{W}_{\mathrm{r},t}} P_{D,t}$. Using the chain rule with $P_{D,t} = Q(z_t)$ where $z_t\! =\! (\!Q^{-1}(P_{FA})\sigma_0 \!- \!\Delta\mu\!)/\!\sigma_1$ from \eqref{eq:pd_final}, and $\phi(\!z) \!=\! \frac{1}{\sqrt{\!2\pi}}e^{\!-z^2/2}$ being the standard normal PDF, we have,
\vspace{-0.05 in}
\begin{equation}
\nabla_{\mathbf{W}_{\mathrm{r},t}} P_{D,t} = -\phi(z_t) \nabla_{\mathbf{W}_{\mathrm{r},t}} z_t.
\vspace{-0.06 in}
\label{eq:pd_gradient_chain}
\end{equation}
where $\nabla_{\!\mathbf{W}_{\mathrm{r},t}} z_t\! \!=\! \frac{1}{\sigma_1}\left(\!Q^{-1}\!(P_{FA})\nabla_{\!\mathbf{W}_{\mathrm{r},t}}\!\sigma_0 \!-\! \nabla_{\!\mathbf{W}_{\mathrm{r},t}}\!\Delta\mu\! 
- \!z_t\nabla_{\!\mathbf{W}_{\mathrm{r},t}}\!\sigma_1\!\right)$. Using the matrix derivative identity $\!\nabla \!\mathbf{A}^{\!-1}\! = \!-\mathbf{A}^{\!-1}\!(\!\nabla \!\mathbf{A}\!)\mathbf{A}^{\!-1}\!$ obtaining $\mathbf{Q}_t$ gradient,
\vspace{-0.05 in}
\begin{equation}
\nabla_{\mathbf{W}_{\mathrm{r},t}} \mathbf{Q}_t \!=\! -\frac{2d_{\mathrm{r},t}^{\zeta}}{L}\mathbf{R}_{1,t}^{-1}\mathbf{H}_t\!\mathbf{W}_{\mathrm{r},t}\!\mathbf{W}_{\mathrm{r},t}^H\mathbf{H}_t^H\!\mathbf{R}_{1,t}^{-1}.
\vspace{-0.06 in}
\label{eq:grad_qt}
\end{equation}
The gradient of $\!\Delta\mu\!$ using the cyclic property of trace and \eqref{eq:grad_qt} is,\!
\vspace{-0.05 in}
\begin{equation}
\nabla_{\mathbf{W}_{\mathrm{r},t}} \Delta\mu \!=\! \frac{2d_{\mathrm{r},t}^{\zeta}}{L}\mathbf{W}_{\mathrm{r},t}^H\mathbf{H}_t^H[\mathbf{Q}_t \!-\! d_{\mathrm{r},t}^{\zeta}\mathbf{R}_{1,t}^{-1}\mathbf{H}_t\mathbf{R}_t\mathbf{H}_t^H\mathbf{R}_{1,t}^{-1}]\mathbf{H}_t.
\vspace{-0.05 in}
\label{eq:grad_deltamu_final}
\end{equation}
For $\sigma_0$ and $\sigma_1$, the gradients follow a similar form. The final gradient can be expressed in matrix form as,
\vspace{-0.05 in}
\begin{equation}
\nabla_{\mathbf{W}_{\mathrm{r},t}} P_{D,t} = -\frac{2d_{\mathrm{r},t}^{\zeta}\phi(z_t)}{L\sigma_1}\mathbf{W}_{\mathrm{r},t}^H\mathbf{H}_t^H\mathbf{M}_t\mathbf{H}_t,
\label{eq:pd_gradient_final}
\vspace{-0.06 in}
\end{equation}
where $\mathbf{M}_t = \mathbf{Q}_t - d_{\mathrm{r},t}^{\zeta}\mathbf{R}_{1,t}^{-1}\mathbf{H}_t\mathbf{R}_t\mathbf{H}_t^H\mathbf{R}_{1,t}^{-1} + \alpha_t\mathbf{Q}_t\mathbf{R}_{1,t}^{-1}\mathbf{Q}_t\mathbf{R}_{1,t}^{-1}$ with $\alpha_t\! = \!(Q^{-1}(\!P_{FA}\!)\sigma_n^2)/(2\sigma_0\sigma_1\!\sqrt{L}) \!- \!z_t\sigma_n^2/\!\sigma_0\!\sqrt{L}$.
\vspace{-0.07 in}
\subsection{Gradients of BER w.r.t $\mathbf{W}_{\mathrm{c}}$}
Using the chain rule, the gradient of $\mathrm{BER}_k$ w.r.t. $\mathbf{W}_\mathrm{c}$ is:
\begin{equation}\vspace{-0.05 in}
\nabla_{\mathbf{W}_\mathrm{c}} \mathrm{BER}_k = \frac{\partial \mathrm{BER}_k}{\partial \mathrm{SINR}_k} \cdot \nabla_{\mathbf{W}_\mathrm{c}} \mathrm{SINR}_k,
\label{ber_chain1}\vspace{-0.06 in}
\end{equation}
where the partial derivative is,
\vspace{-0.05 in}
\begin{equation}
\frac{\partial \mathrm{BER}_k}{\partial \mathrm{SINR}_k} = -\frac{\lambda_\mathrm{c} \sqrt{\beta_\mathrm{c}}}{2 M \log_2 M \sqrt{2\pi \mathrm{SINR}_k}} e^{-\beta_c \mathrm{SINR}_k / 2},\vspace{-0.06 in}
\label{dber_dsinr}
\end{equation}
and $\nabla_{\mathbf{W}_\mathrm{c}} \mathrm{SINR}_k$ is derived. Let $\mathcal{D} = d_{\mathrm{c},k}^{-\zeta} \sum_{i \neq k} |\mathbf{h}_k^T \mathbf{w}_{\mathrm{c},i}|^2 + d_{\mathrm{c},k}^{-\zeta} \|\mathbf{h}_k^T \mathbf{W}_\mathrm{r}\|^2 + \sigma_n^2$. Then,
\vspace{-0.05 in}
\begin{align}
\nabla_{\mathbf{W}_\mathrm{c}} \mathrm{SINR}_k &=
\frac{2 d_{\mathrm{c},k}^{-\zeta}}{\mathcal{D}} \mathbf{h}_k \mathbf{h}_k^T \mathbf{w}_{\mathrm{c},k}\mathbf{e}_k^T \nonumber\\
&-\!\frac{2 d_{\mathrm{c},k}^{-2\zeta}}{\mathcal{D}^2} \|\mathbf{h}_k^T \!\mathbf{w}_{\mathrm{c},k}\|^2 \mathbf{h}_k \mathbf{h}_k^T \mathbf{W}_{\!\mathrm{c}}(\mathbf{I}\!-\!\mathbf{e}_k \mathbf{e}_k^T ),
\label{dsinr_dwc}
\vspace{-0.06 in}
\end{align}
where $\mathbf{e}_k$ $k$-th standard basis vector. Substituting \eqref{dber_dsinr} and \eqref{dsinr_dwc} into \eqref{ber_chain1}, we get the gradient $\nabla_{\mathbf{W}_\mathrm{c}}\mathrm{BER}_k$.
\vspace{-0.1 in}
\bibliographystyle{IEEEtran}
\bibliography{KLD_ISAC_paper_v3.bib}
\end{document}